# Mechano-Bactericidal Surfaces Achieved by Epitaxial Growth of Metal-Organic Frameworks

Zhejian Cao[1,2*], Santosh Pandit[1], Francoise M. Amombo Noa[3,4], Jian Zhang[1], Wengeng Gao[1], Shadi Rahimi[1], Lars Öhrström[3*], and Ivan Mijakovic[1,2,5*]

[1]Department of Life Sciences, Chalmers University of Technology, Gothenburg SE-41296, Sweden.
[2]Wallenberg Initiative Materials Science for Sustainability, Department of Life Sciences, Chalmers University of Technology, Gothenburg SE-41296, Sweden.
[3]Department of Chemistry and Chemical Engineering, Chalmers University of Technology, Gothenburg SE-41296, Sweden
[4]Ecole des Sciences de la Santé, Université Catholique D'Afrique Centrale, B.P.1110 Yaoundé, Cameroun
[5]The Novo Nordisk Foundation, Center for Biosustainability, Technical University of Denmark, DK-2800 Kogens Lyngby, Denmark

E-mail: zhejian@chalmers.se, ohrstrom@chalmers.se, ivan.mijakovic@chalmers.se

Abstract
Mechano-bactericidal (MB) surfaces have been proposed as an emerging strategy for preventing biofilm formation. Unlike antibiotics and metal ions that chemically interfere with cellular processes, MB nanostructures cause physical damage to the bacteria. The antibacterial performance of artificial MB surfaces relies on rational control of surface features, which is difficult to achieve for large surfaces in real-life applications. Herein, we report a facile and scalable method for fabricating MB surfaces based on metal-organic frameworks (MOFs) using epitaxial MOF-on-MOF hybrids as building blocks with nanopillars of less than 5 nm tip diameter, 200 nm base diameter, and 300 nm length. Two methods of MOF surface assembly, *in-situ* growth and *ex-situ* dropcasting, result in surfaces with nanopillars in different orientations, both presenting MB actions (bactericidal efficiency of 83% for *E. coli*). Distinct MB mechanisms, including stretching, impaling, and apoptosis-like death induced by mechanical injury are discussed with the observed bacterial morphology on the obtained MOF surfaces.

## 1. Introduction
Bacterial biofilms are the linchpin of persistent infections and biofouling[1]. In particular, medical indwelling devices, dental devices, and prostheses provide accessible surfaces conducive to biofilm development, contributing to approximately 80% of chronic and nosocomial infections[2]. Mitigation strategies based on chemical interference with cellular processes have only limited success as biofilm cells possess much higher resistance to antibiotics and the human immune system compared to their planktonic counterparts[3]. Furthermore, misuse of antibiotics has accelerated the spread of antimicrobial resistance (AMR), which is currently considered one of the largest threats to global health[4]. Biofilm formation starts with bacterial attachment on surfaces, after which bacteria proliferate and adhere to each other within a self-produced extracellular polymeric matrix[5]. Therefore, preventing the initial attachment of bacteria to a surface could be an effective mitigation strategy to slow down or even preclude the formation of mature biofilms[6,7].

Mechano-bactericidal (MB) surfaces have emerged as a material-centric solution for preventing biofilm formation[8]. Unlike antibiotics or metal ions that attack bacteria chemically, MB actions involve nanostructures inducing rupture and death of bacterial cells through physicomechanical interactions[9]. MB actions can be found in natural and artificial nanostructured surfaces (Supplementary Section 1), such as the cicada wings[10], gecko skin[11], black silicon[12], and vertical graphene[13]. In addition to their intrinsic bactericidal efficacy, nanostructured surfaces are reported to promote the effectiveness of antibiotics, which can potentially reduce antibiotic dosage[14]. Nevertheless, real-life applications of artificial MB surfaces are limited by complex fabrication processes, as nanostructures are required to provide MB efficacy[15,16]. For instance, we have previously demonstrated that optimal bacterial killing was achieved with graphene nanosheets aligned perpendicular to the surface, not thicker than 10 atom layers, and spaced so that bacterial cells cannot fit between adjacent spikes[17]. However, the fabrication of such surfaces is typically expensive, laborious, and requires sophisticated equipment[18]. These critical challenges impede the scalability and integration of MB surfaces into practical applications. We propose that metal-organic frameworks (MOFs), with their access to various morphologies, including sharp-point shape crystal growth, could be used to create MB surfaces using a facile and economical fabrication process with the scalability required for real-life applications.

MOFs are emerging porous materials with crystalline structures, designable geometry, tailorable chemical composition, and mild synthesis conditions[19,20]. Despite being relatively novel materials, the ton-scale production of many MOFs has been reported[21–23], and size and morphology control are routine in industrial crystallization[24]. Furthermore, MOFs are currently developed, academically and in start-up companies, for various bio-applications, such as drug delivery, enzyme immobilization, and biosensors[25,26]. Therefore, MOFs could be a promising candidate as building blocks for scalable MB surfaces. In this study, we control the orientation of MOFs based on biocompatible metal ions to create spike-like surfaces with MB properties. This is a different approach from many MOFs that have been reported as potential antibacterial agents due to their ability to release toxic metal ions (Supplementary Section 2), including silver (Ag), copper (Cu), zinc (Zn), and cobalt (Co) ions[27,28]. Here we report a facile process to fabricate MOF MB surfaces through epitaxial growth of MOF-on-MOF hybrids with sharp-tip features, using two biocompatible MOFs, MIL-88B(Fe) and UiO-66(Zr)[29,30]. These MOF particles presented a caltrop-like three-dimensional (3D) structure and were assembled on surfaces through *in-situ* growth and *ex-situ* dropcasting methods. The obtained MOF surfaces presented MB effects towards both Gram-positive and Gram-negative bacteria. Our approach and results cast light on the obstacles to achieving MB surfaces with rational-controlled geometric features in a facile manner and open a novel sight for applying MOFs for antibacterial applications.

## 2. Rational control of geometric features in MOF-on-MOF surfaces

The size, orientation, and density of nanopatterns are critical for the performance of natural and artificial MB surfaces[31]. MIL-88B(Fe) was selected as the nanopillar due to its nanoscale size, spike-like geometry, and low cytotoxicity[32,33]. However, arranging MIL-88B nanopillars perpendicularly to be accessible to bacteria is challenging. Furthermore, MIL-88B as nanopillars would have to be precisely spaced to avoid the bed-of-nails effect[34] and intercalation of bacteria in between adjacent pillars[35,36]. Based on these criteria, we applied a core-satellite MOF-on-MOF strategy, where UiO-66 was used as the core and MIL-88B as the satellite[37,38]. The caltrop-like MIL-88B-on-UiO-66 (denoted as MoU) hybrids were assembled with two approaches, i.e., *in-situ* growth and *ex-situ* dropcasting, to achieve MOF MB surfaces shown in Fig. 1. The *in-situ* growth approach (Fig. 1a) resulted in a one-pin up orientation (Fig. 1b, Supplementary Video 1), where a UiO-66 layer first *in-situ* grew on the substrate, followed by the epitaxial growth of the MIL-88B onto the UiO-66 crystals. The length of the MIL-88B nanopillars was around 300 nm and the base and tip diameter of the nanopillars were around 200 nm and less than 5 nm, respectively, as shown in the transmission electron microscopy (TEM) image (Fig. 1e). The pitch distance between the tips of vertical MIL-88B nanopillars was around 500 nm, which was controlled by the distance between the centers of UiO-66 cores (Supplementary Section 3), as demonstrated in the scanning electron microscopy (SEM) images (Fig. 1cd). This position correlation between the nanopillars and the core MOF opens the possibility to rationally adjust the surface features in the MOF MB surfaces, as the distance between adjacent UiO-66 particles and the length of the MIL-88B can be modified (Supplementary Section 4)[39–42]. The other method we used was *ex-situ* dropcasting. In this approach, MoU hybrids were synthesized and then dropcast to substrates as illustrated in Fig. 1f.  When dropcast, due to the unique eight-pin 3D structure of the MoU hybrid, MoU tended to land with four nanopillars on the surface to achieve mechanical equilibrium, resulting in a four-pin up orientation (Fig. 1g, Supplementary Video 2), as demonstrated in the TEM image (Fig. 1k). Notably, to achieve a fully covered surface, some areas of the dropcast MOF coating consisted of multilayered MoU, which could lead to a random orientation and distance of the MIL-88B nanopillars (Fig. 1j). Dropcast UiO-66 (Fig. 1h) and MiL-88B (Fig. 1i) surfaces lacked the critical geometric features of MB surfaces compared with MoU surfaces, including sharp nanostructures and vertical orientation. Both UiO-66[43,44] and MIL-88B[45] are reported to be suitable for large-scale production and are commercially available on the market[46–48]. Since MoU synthesis is a combination of UiO-66 and MIL-88B production, the resulting material can be considered both scalable and economical. Furthermore, our MoU surface assembly methods required no sophisticated equipment and involved low fabrication temperatures, i.e., 120 ºC for the MOF synthesis by *in-situ* growth and room temperature (around 20 ºC) by *ex-situ* dropcasting. This allows for large-scale MOF MB surfaces with a wide range of feasible substrates, especially materials that cannot stand high temperatures (Supplementary Section 5).

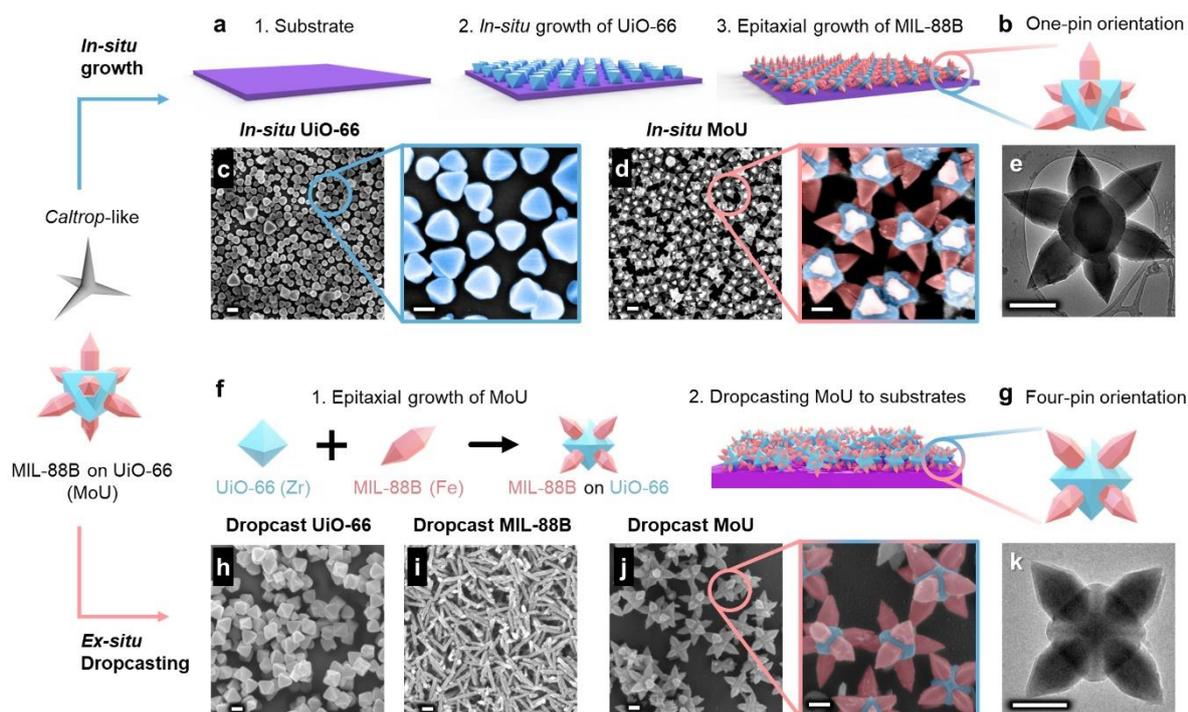

Fig. 1 **MOF mechano-bactericidal (MB) surfaces.** Caltrop-like MIL-88B-on-UiO-66 (MoU) hybrids used as the building blocks for MB surfaces through two assembly methods. (a) Schematic showing MoU surfaces through *in-situ* growth: 1. substrate, 2. *in-situ* growth of UiO-66 on the substrate, 3. epitaxial growth of MIL-88B-on-UiO-66 (MoU); (b) Schematic illustration of the one-pin up orientation in *in-situ* MoU surfaces; SEM images of (c) *in-situ* UiO-66 surface and (d) *in-situ* MoU surface after epitaxial growth of MIL-88B; (e) TEM image of MoU hybrid with one-pin up orientation; (f) Schematic showing MoU surfaces through *ex-situ* dropcasting, 1. MoU hybrids through epitaxial growth, 2. Dropcasting MoU to the substrate; (g) Schematic illustration of the four-pin up orientation in dropcast MoU surfaces; SEM images of dropcast (h) UiO-66, (i) MIL-88B, and (j) MoU, zoomed-in SEM image false-colored with UiO-66 in blue and MIL-88B in pink; (k) TEM image of MoU hybrid with a four-pin up orientation. Scale bar: 200 nm

To verify their crystal structure and chemical composition, the obtained surfaces were characterized by powder X-ray diffraction (XRD) and X-ray photoelectron spectroscopy (XPS). The obtained UiO-66 and MIL-88B matched the characteristic peaks of the simulated XRD patterns (Fig. 2a, Supplementary Section 6), revealing successful MOF synthesis and good crystallinity. Furthermore, *in-situ* and dropcast MoU contained characteristic peaks of UiO-66 and MIL-88B, indicating a successful epitaxial MOF-on-MOF growth. The XPS survey spectra in Fig. 3b show that C, O, Fe, and Zr elements were found on the MoU surfaces, where Fe is from the MIL-88B nanopillars and Zr from UiO-66 cores. High-resolution XPS spectra for Fe *2p* and Zr *3d* confirm the existence of $Fe^{3+}$ and $Zr^{4+}$ in the obtained MoU (Supplementary Section 7). SEM/TEM images (Fig. 1), XRD and XPS results (Fig. 2ab) jointly confirm the success of MOF-on-MOF synthesis in obtaining MoU surfaces, where the nanopillar structures are MIL-88B(Fe) and the core structures are UiO-66(Zr).

The wettability of surfaces may influence bacterial adhesion[8,49]. Our *in-situ* MoU surface had a water contact angle (CA) of 127°, compared to the 55° of the intact Si substrates (Fig. 2c), indicating increased hydrophobicity. The other MOF surfaces we fabricated were also hydrophobic (Supplementary Section 8). The enhancement of the hydrophobicity can likely be attributed to the hydrophobicity of MOFs[50] and the nanostructure-induced surface roughness[51].

Atomic force microscopy (AFM) was utilized for mapping the topography of the MOF surfaces. The AFM mapping (Fig. 2d-f) confirmed that MOFs as building blocks created spaced nanopatterns. Considering that most parts of the caltrop-like MOF-on-MOF structures are not perpendicular to the substrates, topographical parameters obtained by AFM scanning, such as surface roughness, may not reveal the actual surface features which are crucial for MB properties of the obtained MOF surfaces. Therefore, we mainly used AFM in this work to understand the orientation differences in the *in-situ* and *ex-situ* MoU surfaces by investigating the zoomed-in cross-section profiles for individual MOF structures. As demonstrated in Fig. 2d, the *in-situ* grown UiO-66 provided a near-horizontal plane for the MIL-88B to vertically grow epitaxially. Therefore, the MIL-88B nanopillars in the *in-situ* MoU surface presented a perpendicular orientation to the substrate (Fig. 2e), resulting in one-pin up orientation. By contrast, the MIL-88B nanopillars issued from the dropcast coating tended to be at approximately 35º to the substrate, leading to a four-pin up orientation. Moreover, due to multilayer stacking, nanopillars in dropcast MoU presented a random orientation (Fig. 2f). These findings confirmed that the orientation of the MOF nanopillars can be arranged at different angles by *in-situ* and *ex-situ* methods.

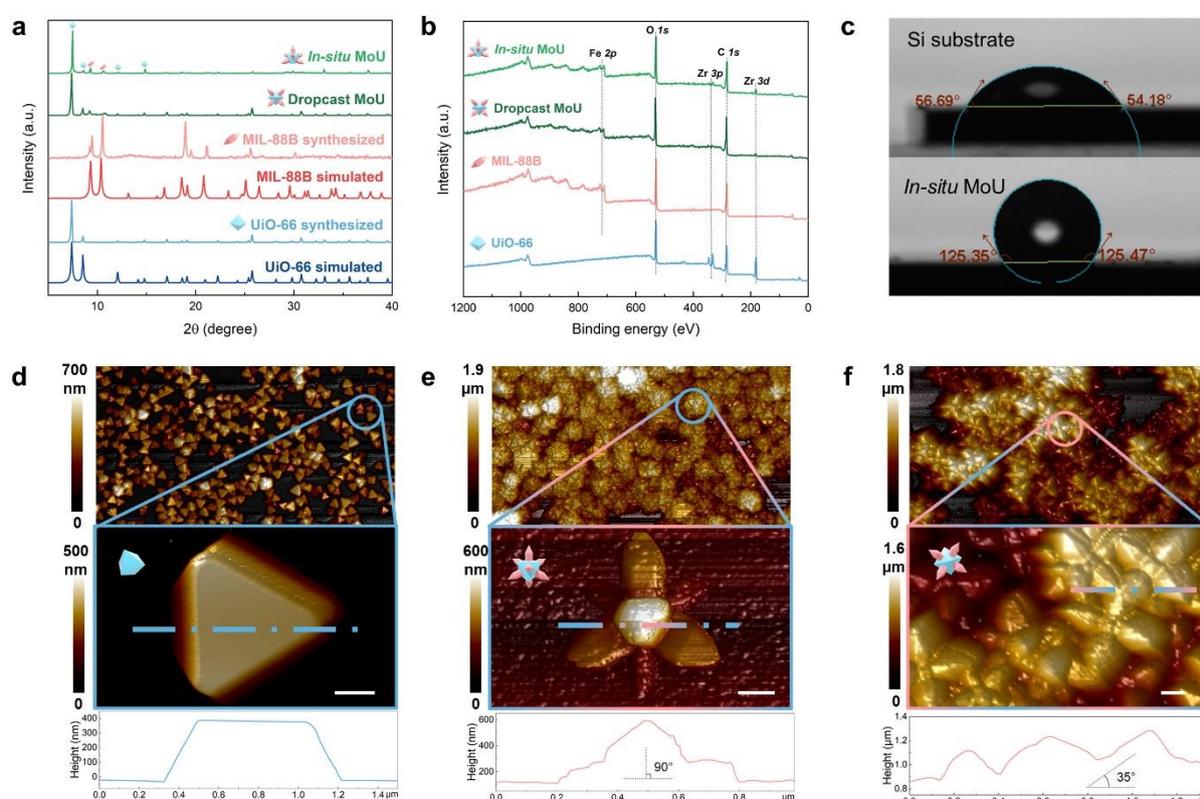

Fig. 2 **Material characterization of MOF MB surfaces.** (a) XRD patterns of *in-situ* and dropcast MoU surfaces, UiO-66, MIL-88B, and their simulated patterns. The characteristic peaks of UiO-66 (blue-octahedron) and MIL-88B (pink-rod) are marked at corresponding positions; (b) XPS survey spectra of the obtained MOF surfaces with C, O, Fe, and Zr characteristic peaks marked; (c) The water contact angle of the silicon substrate and the *in-situ* MoU surfaces showing the enhancement of hydrophobicity after MOF coating. The AFM scanning and cross-section profile of (d) *in-situ* UiO-66, providing a horizontal plane for MIL-88B epitaxial growth (e) *in-situ* MoU, showing one-pin up orientation with a near perpendicular MIL-88B nanopillar, and (f) dropcast MoU surfaces, showing four-pin up orientation with random angles of the MIL-88B nanopillars, scale bar: 200 nm

## 3. Mechano-bactericidal actions of MOF surfaces

MB efficacy of the obtained MOF surfaces was examined by plate counting of colony forming unit (CFU) and live/dead fluorescent staining, using *Escherichia coli* (*E. coli*) and S*taphylococcus epidermidis* (*S. epidermidis*) as model organisms for Gram-negative and Gram-positive bacteria, respectively. Additionally, multidrug-resistant (MDR) *Staphylococcus aureus* (*S. aureus*) was tested as a top-priority pathogen. Ruling out the chemical cytotoxicity of the MOFs is crucial to the quantitative study of the mechano-bactericidal performance. Therefore, the zone of inhibition test was carried out for all obtained MOF surfaces. No inhibition zone was found in the MOF surfaces, indicating low chemical cytotoxicity of the MOFs (Supplementary Section 9). However, during the tests with liquid culture media, we observed that MIL-88B(Fe) presented bactericidal effects towards the Gram-positive bacteria (both *S. epidermidis* and *S. aureus*). This is most likely caused by Fenton-like reactions from the iron (Fe) ions released from the MIL-88B, of which Gram-positive bacteria are known to be sensitive.[52] This intrinsic bactericidal effect of MIL-88B toward Gram-positive bacteria impeded the quantitative study of the MB performance of the MOF surfaces. Cumulative effects of chemical toxicity and MB effects on Gram-positive bacteria are shown in Supplementary Section 10. This left us with *E. coli* as the model for investigating MOF MB effects. Both the *in-situ* MoU and dropcasting MoU surfaces demonstrated MB performance according to the CFU bar chart of the attached *E. coli* after 24h growth (Fig. 3a). Notably, individual UiO-66, MIL-88B, and physical mixture of UiO-66 + MIL-88B (denoted as U+M) did not show significant bactericidal efficiency according to their CFU results, which confirmed no/negligible chemical killing effect was involved for *E. coli*. For the *in-situ* MoU surfaces, the bactericidal efficiency was relatively low, at around 32%. While the bactericidal efficiency of the dropcast MoU surfaces reached around 83%, landing in the upper tier (over 80%) of the reported solely MB active surfaces (Supplementary Section 1). This can probably be explained by the fact that there were fewer gap areas on the dropcast MoU surfaces (15%) than on the *in-situ* MoU surfaces (35%), as analyzed by our previously reported image analysis methods (Fig. 3c, Supplementary Section 11)[35,53]. The gap areas may not be effectively protected, as bacteria could attach and survive. This could also be revealed by the live/dead staining images in Fig. 3b, where green signals indicate live bacteria and red are dead bacteria. *E. coli* formed a dense biofilm on the control surface, while several isolated bacterial clusters were observed to attach on the *in-situ* MoU surface. For the dropcast MoU surface, separated individual bacteria were observed in the live/dead staining images (Fig. 3b), corresponding to a much smaller unprotected area in dropcast surfaces seen in Fig. 3c2. The bactericidal efficiency of the dropcast MoU surface dropped from 83% for 24h growth to 51% for 72h growth (Supplementary Section 12), which could result from the coverage of the nanostructures by the debris, such as the killed bacteria[54,55]. Therefore, an efficient surface cleaning strategy to remove debris could be essential for the long-term protection and reusability of the MOF MB surfaces[56,57].

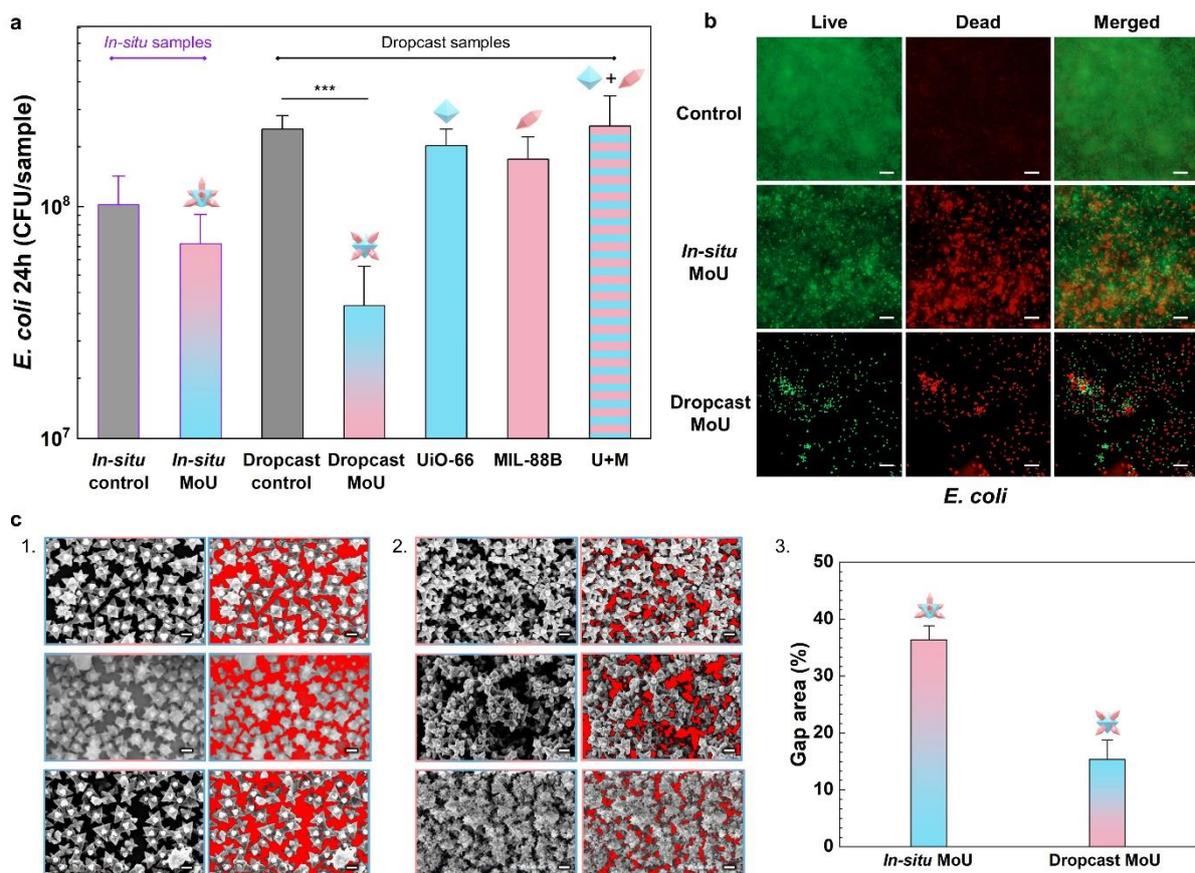

Fig. 3 **Antibacterial performance of MOF MB surfaces.** (a) CFU counting results of attached *E. coli* with 24h growth on *in-situ* MoU surfaces and dropcasting surfaces, including MoU, UiO-66, MIL-88B, and UiO-66 + MIL-88B (U+M). Data represent the mean ± standard deviation of three biological replicates (***p < 0.001). (b) The live/dead fluorescent staining images of attached *E. coli* with 24h growth on *in-situ* and dropcasting MoU surfaces, green indicating live bacteria and red indicating dead bacteria, scale bar: 10 μm. (c) SEM images of (1) *in-situ* MoU and (2) dropcast MoU surfaces with gap areas marked in red, scale bar: 200 nm. (3) The bar chart of the gap area percentage of *in-situ* and dropcast MoU surfaces based on three different regions

To investigate the inactivation mechanisms of the obtained MOF surfaces, the morphology of the bacteria exposed to MOFs was characterized with SEM. As seen in Fig. 4ab, *E. coli* and *S. epidermidis* grew dense biofilms on the control glass surfaces. On the contrary, neither of the bacteria was able to develop a continuous biofilm and both presented ruptured structures on *in-situ* and dropcast MoU surfaces. Different types of bacterial deformation were observed on the two different MoU surfaces. For instance, cell rupture by *stretching* was found in *E. coli* on *in-situ* MoU surfaces (Fig. 4a5), where the *E. coli* cell was pinned by several MOF nanopillars, and the bacterial envelope was deformed and lost their spherocylinder shape by stretching and tearing[7]. Cell rupture by direct *impaling* was found in *E. coli* on dropcast MoU surfaces (Fig. 4a6), where MOF nanopillars directly penetrated the bacterial membrane, resulting in deflated morphology and cytoplasm leakage[58]. Direct impaling was also observed in *S. epidermidis*, illustrated by triangle-like holes found on cells exposed to *in-situ* MoU surfaces (Fig. 4b5). Apart from stretching and impaling, some bacteria suffered non-piercing mechanical injury, e.g., *S. epidermidis* on the dropcast MoU surfaces presented a squeezed morphology (Fig. 4b6). Even though the MOF nanopillars did not lead to direct penetration, the mechanical stress alone is known to induce oxidative stress ultimately leading to apoptosis-like death[59,60]. Tilted SEM images were acquired to provide close-up views of the interactions between bacterial cells and MOF structures, where the interfaces of the nanopillars impaling bacterial envelopes were revealed (Supplementary Section 13).

Similar rupture phenomena were found in MDR *S. aureus* (Supplementary Section 14). Notably, both *in-situ* and *ex-situ* MoU surfaces presented intact structures and surface coverage after antibacterial evaluation, suggesting good structural stability (Supplementary Section 15). Understanding the stress distribution on the bacterial cell is essential for understanding the killing mechanism of MB surfaces. Therefore, we conducted a simulation of stress analysis for *E. coli* and *S. aureus* on *in-situ* MoU surfaces (Supplementary Section 16). As shown in the stress contour in Fig. 4cd, the maximum stress on the bacterial envelope is 114 MPa for *E. coli* and 54 MPa for *S. aureus*. These maxima exceed the critical elastic stress of the bacteria[61]. Since the maximum stress occurs at the areas contacting tips of the nanopillars, the simulation results support the feasibility of the MOF nanopillars puncturing the bacterial membrane, as illustrated in Fig. 4e (Supplementary Video 3). Two SEM images of MoU directly impaling *E. coli* and *S. epidermidis* with the tip of the nanopillar puncturing bacterial envelope are shown in Fig. 4f.

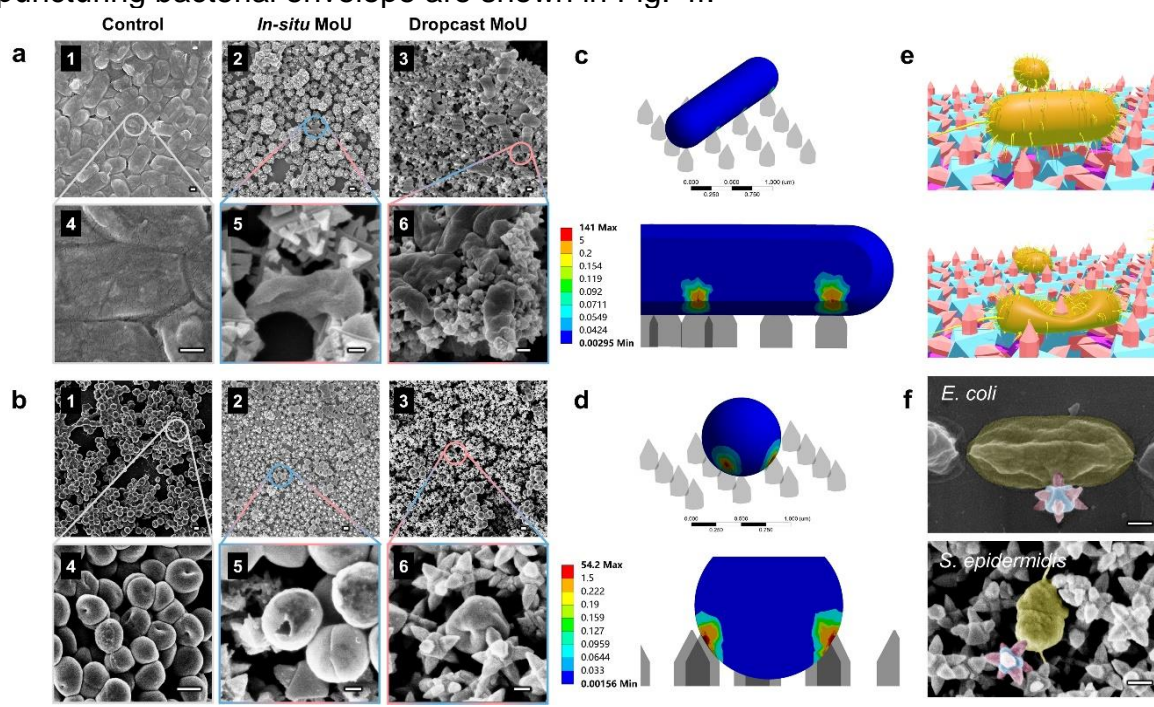

Fig. 4 **Mechanism study of the MOF MB surfaces.** SEM images of attached bacteria on control, *in-situ* MoU, and dropcast MoU surfaces with 24h growth: (a) *E. coli*; (b) *S. epidermidis*. Stress contour (unit: MPa) of bacteria on *in-situ* MoU surfaces: (c) *E. coli*; (d) *S. aureus.* (e) Illustration of bacterial rupture by MoU surfaces. (f) SEM images of MoU impaling *E. coli* and *S. epidermidis*, false-colored with bacteria in yellow, UiO-66 in blue, and MIL-88B in pink. Scale bar: 200 nm

## 4. Conclusion

In this work, we designed mechano-bactericidal (MB) surfaces using the epitaxial MOF-on-MOF hybrids. The caltrop-like MIL-88B-on-UiO-66 (MoU) hybrids were assembled on the surfaces using two approaches: *in-situ* growth and *ex-situ* dropcasting, resulting in one-pin and four-pin up orientations of the MoU, respectively. We demonstrated that the features of the MoU surfaces can be rationally controlled to fulfill the key criteria for MB actions: sharp nanostructures with vertical orientation and adequate inter-spacing. Three types of MB actions were observed, including stretching, impaling, and apoptosis-like death induced by mechanical injury. Due to low fabrication temperature, a wide range of feasible substrates, the waiver of sophisticated equipment, and the scalability of our MOF-on-MOF MB surfaces, we believe that they could make a substantial contribution to the mitigation of the initial attachment of bacteria and slowing down biofilm formation in various applications, such as food packaging, and protection of surfaces on medical implants and devices.

## 5. Method

*Fabrication of the MOF MB Surfaces*: MOFs including UiO-66, MIL-88B, and MoU were obtained using solvothermal synthesis by modifying reported processes[37,38,40]. The *in-situ* UiO-66 surfaces were obtained by loading the Si chips (6 mm × 6 mm) into the autoclave during UiO-66 synthesis. *In-situ* MoU surfaces were obtained by loading the *in-situ* UiO-66 surfaces into the Pyrex tube during MoU synthesis. Dropcast MOF surfaces were obtained by dropcasting 50 μL MOF solution (5 mg mL$^{-1}$ in ethanol) on the round glass slide (diameter 10 mm). All the samples were dried in a static vacuum oven at 60 °C overnight before tests. More details on the MOF synthesis and fabrication of the MOF MB surfaces are described in Supplementary Section 17.

*Scanning Electron Microscopy (SEM)*: The microstructure of the samples was characterized by SEM (Zeiss Supra 55VP) with 5 kV accelerating voltage. The samples containing bacteria were fixed with 3% glutaraldehyde for 2 h and then dehydrated by a series of ethanol concentrations (40%, 50%, 60%, 70%, 80%, and 90%) for 10 min each, and with absolute ethanol for 15 min. All samples were dried in a static vacuum oven for 3 h at 60 °C and coated with a 15 nm gold layer to avoid charging.

*Transmission Electron Microscopy (TEM)*: The TEM images were acquired from an FEI Tecnai T20 microscope, equipped with a LaB6 filament and operating at 200 kV accelerating voltage. The MOF solution (1 mg mL$^{-1}$ in ethanol) was dropcast on carbon-film supported copper grids (Sigma-Aldrich) and dried in a static vacuum oven for 3 h at 60 °C before tests.

*X-Ray Diffraction (XRD)*: The powder X-ray diffraction (PXRD) was characterized by using a Bruker XRD D8 Advance with Cu Kα X-ray source (λ = 1.54 Å) at room temperature, with a scanning range 2θ from 5° to 40°. Simulated PXRD patterns were calculated with Mercury software with crystal data from Cambridge Structure Database (CSD).

*X-Ray Photoelectron Spectroscopy (XPS)*: The surface chemistry of the obtained MOF samples was studied with a PHI VersaProbe III X-ray photoelectron spectroscopy (XPS) instrument with Al Kα X-ray source. The data analysis was performed with CasaXPS software.

*Contact Angle (CA) Measurements*: The wettability of the obtained surfaces was characterized by measuring the water contact angle in the air using an optical tensiometer (Attension, Biolin Scientific). The images were taken within 5 seconds of the droplets being dispended on the surfaces. Statistical averaging of 3 replicates was performed on each sample.

*Atomic Force Microscopy (AFM)*: The surface topography was studied using atomic force microscopy using Dimension ICON (Bruker) with tapping mode with Bruker RTESP-300 AFM probes. All samples were coated with a 15 nm gold layer to reduce the risk of picking MOF particles during AFM measurement.

*Computational Stress Analysis*: To study the stress of the bacterial envelope on the MOF surfaces, we used commercial software (ANSYS 2024 R1) as a finite element analysis environment. The shape of *E. coli* and *S. aureus* bacteria was modeled as spherocylinder and sphere, respectively. The geometry parameters of the bacterial size and MOF size were obtained from the SEM and TEM images. Other parameters including Young's modulus of the bacterial envelope and the adhesion forces between bacteria and substrates were collected from reported literature. All parameters used for the simulation were summarized in Supplementary Section 16.

*Evaluation of Bacterial Viability*: The antibacterial performance was evaluated through plate counting of colony-forming unit (CFU) method as reported in our previous studies[17,36]. The bacterial strains *Escherichia coli* (UTI89), *Staphylococcus epidermidis* (ATCC 35984), and *Staphylococcus aureus* (CCUG 35571) were obtained from Gothenburg University Culture Collection (CCUG) and used for antibacterial evaluation on the obtained surfaces. Single colonies of each bacterial strain were grown in a liquid medium (5mL), Luria-Bertani (LB) broth for *E. coli* and tryptic soy broth (TSB) for *S. epidermidis* and *S. aureus*, at 37°C overnight. Then, 25 µL overnight bacterial culture was added to 5 mL fresh medium to obtain an inoculum containing $2 - 5 \times 10^6$ CFU mL$^{-1}$ bacteria, where the bacterial cell density was verified by plate counting. The inoculum was then loaded onto the tested surfaces, at 40 µL and 100 µL to the silicon substrated (square, 6 mm × 6 mm) samples and glass slide substrated (round, diameter 10 mm) samples, respectively, to make sure the MOF surfaces fully covered while not contaminating the uncoated surfaces. Notably, to avoid the evaporation of the culture media on the sample surfaces, tested samples were placed in the middle part of a 24-well plate, while the surrounding empty wells at the edges as well as the gaps of the plate were filled with sterilized distilled water. After 24 h bacterial growth at 37 °C, the media culture was removed from the surface and irreversibly attached bacteria were detached from surfaces by sonication (Digital Sonifier, Branson, 10% amplitude, 30 s) and collected in saline solution (5 mL of 0.89% NaCl). Thereafter, the collected bacteria were diluted (×10) serially and plated in agar plates. The plates were incubated at 37 °C for 24h. The number of colonies on the collected plates was counted and the number of viable bacteria (CFU sample$^{-1}$) was then estimated by the number of colonies counted in plates and their corresponding dilution factors. The bactericidal efficiency was obtained by normalizing the CFU counts of each surface corresponding to that of the control surface. The experiments were conducted with three biological replicates and the mean values ± standard deviation are reported. The statistical significance between MoU surface and control surface was examined using independent t-test at $P \leq 0.05$.

*Fluorescence Microscopy*: The live/dead assay was done with fluorescence microscopy. The attached bacteria on the surface were stained with LIVE/DEAD BacLight bacteria viability stains kit L7012 (Invitrogen, Molecular Probes, Inc. Eugene, OR, USA). The kit consists of the green-fluorescent nucleic acid stain SYTO 9 and the red-fluorescent nucleic acid stain propidium iodide (PI). The green-fluorescent dye (SYTO 9) crosses all bacterial membranes and binds to the DNA of bacterial cells. The red-fluorescent PI only crosses damaged bacterial membranes (dead bacteria). Fluorescence microscopy imaging of the attached bacteria was acquired using a Zeiss fluorescence microscope (LeicaCTR4000), after staining the samples with a mixture of SYTO 9 and PI for 20 min. Experiments were performed in three biological replicates and representative images are presented.

**Acknowledgments**
This work was supported by grants from Wallenberg Initiative Materials Science for Sustainability (WISE) funded by the Knut and Alice Wallenberg Foundation (KAW), Chalmers Area of Advance Nano, NordForsk (Project No. 105121), Novo Nordisk Foundation (NNF20CC0035580), and the Independent Research Fund Denmark (DFF 3164-00026B) to IM, and Vetenskapsrådet (2020-04096) to SP. ZC acknowledges the lab assistance from Nihal Kottan and Gan Wang. The authors acknowledge Chalmers research infrastructures, including applied chemistry, CMAL, and MC2 for providing the training and testing equipment.

**Declaration of Competing Interest**
The authors declare that they have no known competing financial interests or personal relationships that could have appeared to influence the work reported in this paper.

**Author contributions**
ZC: Conceptualization, Methodology, Formal analysis, Investigation, Data Curation, Writing - Original Draft, Writing - Review & Editing, Visualization, Funding acquisition. SP: Methodology, Formal analysis, Investigation, Data Curation, Writing - Review & Editing, Supervision, Funding acquisition. FMAN: Methodology, Investigation, Writing - Review & Editing, Supervision, Funding acquisition. JZ: Methodology, Formal analysis, Data Curation, Writing - Review & Editing. WG: Methodology, Formal analysis, Data Curation, Writing - Review & Editing. SR: Methodology, Writing - Review & Editing. LÖ: Conceptualization, Methodology, Formal analysis, Investigation, Data Curation, Writing - Review & Editing, Supervision, Project administration, Funding acquisition. IM: Conceptualization, Methodology, Formal analysis, Investigation, Data Curation, Writing - Review & Editing, Supervision, Project administration, Funding acquisition.

**Data availability**
Data will be made available on request.

**Reference**

1. Costerton, J. W., Stewart, P. S. & Greenberg, E. P. Bacterial Biofilms: A Common Cause of Persistent Infections. *Science* **284**, 1318–1322 (1999).
2. Choi, V., Rohn, J. L., Stoodley, P., Carugo, D. & Stride, E. Drug delivery strategies for antibiofilm therapy. *Nat. Rev. Microbiol.* 1–18 (2023) doi:10.1038/s41579-023-00905-2.
3. Macek, B. *et al.* Protein post-translational modifications in bacteria. *Nat. Rev. Microbiol.* **17**, 651–664 (2019).
4. Micoli, F., Bagnoli, F., Rappuoli, R. & Serruto, D. The role of vaccines in combatting antimicrobial resistance. *Nat. Rev. Microbiol.* **19**, 287–302 (2021).
5. Vlamakis, H., Chai, Y., Beauregard, P., Losick, R. & Kolter, R. Sticking together: building a biofilm the Bacillus subtilis way. *Nat. Rev. Microbiol.* **11**, 157–168 (2013).
6. Cheng, Y., Feng, G. & Moraru, C. I. Micro- and Nanotopography Sensitive Bacterial Attachment Mechanisms: A Review. *Front. Microbiol.* **10**, (2019).
7. Cheng, Y., Ma, X., Franklin, T., Yang, R. & Moraru, C. I. Mechano-Bactericidal Surfaces: Mechanisms, Nanofabrication, and Prospects for Food Applications. *Annu. Rev. Food Sci. Technol.* **14**, 449–472 (2023).

8. Linklater, D. P. *et al.* Mechano-bactericidal actions of nanostructured surfaces. *Nat. Rev. Microbiol.* **19**, 8–22 (2021).
9. Linklater, D. P. & Ivanova, E. P. Nanostructured antibacterial surfaces – What can be achieved? *Nano Today* **43**, 101404 (2022).
10. Ivanova, E. P. *et al.* Natural Bactericidal Surfaces: Mechanical Rupture of Pseudomonas aeruginosa Cells by Cicada Wings. *Small* **8**, 2489–2494 (2012).
11. Watson, G. S. *et al.* A gecko skin micro/nano structure – A low adhesion, superhydrophobic, anti-wetting, self-cleaning, biocompatible, antibacterial surface. *Acta Biomater.* **21**, 109–122 (2015).
12. Ivanova, E. P. *et al.* Bactericidal activity of black silicon. *Nat. Commun.* **4**, 2838 (2013).
13. Pham, V. T. H. *et al.* Graphene Induces Formation of Pores That Kill Spherical and Rod-Shaped Bacteria. *ACS Nano* **9**, 8458–8467 (2015).
14. Bright, R. *et al.* Surfaces Containing Sharp Nanostructures Enhance Antibiotic Efficacy. *Nano Lett.* **22**, 6724–6731 (2022).
15. Liu, T. *et al.* Mechanism Study of Bacteria Killed on Nanostructures. *J. Phys. Chem. B* **123**, 8686–8696 (2019).
16. Ghai, V. *et al.* Achieving Long-Range Arbitrary Uniform Alignment of Nanostructures in Magnetic Fields. *Adv. Funct. Mater.* 2406875 (2024) doi:10.1002/adfm.202406875.
17. Pandit, S. *et al.* Vertically Aligned Graphene Coating is Bactericidal and Prevents the Formation of Bacterial Biofilms. *Adv. Mater. Interfaces* **5**, 1701331 (2018).
18. Wu, S., Zuber, F., Maniura-Weber, K., Brugger, J. & Ren, Q. Nanostructured surface topographies have an effect on bactericidal activity. *J. Nanobiotechnology* **16**, 20 (2018).
19. Öhrström, L. & Amombo Noa, F. M. *Metal-Organic Frameworks*. (American Chemical Society, 2020). doi:10.1021/acs.infocus.7e4004.
20. Li, H., Eddaoudi, M., O'Keeffe, M. & Yaghi, O. M. Design and synthesis of an exceptionally stable and highly porous metal-organic framework. *Nature* **402**, 276–279 (1999).
21. Chen, Z. *et al.* The state of the field: from inception to commercialization of metal–organic frameworks. *Faraday Discuss.* **225**, 9–69 (2021).
22. Lin, J.-B. *et al.* A scalable metal-organic framework as a durable physisorbent for carbon dioxide capture. *Science* **374**, 1464–1469 (2021).
23. Zheng, Z. *et al.* High-yield, green and scalable methods for producing MOF-303 for water harvesting from desert air. *Nat. Protoc.* **18**, 136–156 (2023).
24. Myerson, A. *Handbook of Industrial Crystallization*. (Butterworth-Heinemann, 2002).
25. Giménez-Marqués, M., Hidalgo, T., Serre, C. & Horcajada, P. Nanostructured metal–organic frameworks and their bio-related applications. *Coord. Chem. Rev.* **307**, 342–360 (2016).
26. Sun, Y. *et al.* Metal–Organic Framework Nanocarriers for Drug Delivery in Biomedical Applications. *Nano-Micro Lett.* **12**, 103 (2020).
27. Wyszogrodzka, G., Marszałek, B., Gil, B. & Dorożyński, P. Metal-organic frameworks: mechanisms of antibacterial action and potential applications. *Drug Discov. Today* **21**, 1009–1018 (2016).
28. Shen, M. *et al.* Antibacterial applications of metal–organic frameworks and their composites. *Compr. Rev. Food Sci. Food Saf.* **19**, 1397–1419 (2020).
29. Ruyra, À. *et al.* Synthesis, Culture Medium Stability, and In Vitro and In Vivo Zebrafish Embryo Toxicity of Metal–Organic Framework Nanoparticles. *Chem. – Eur. J.* **21**, 2508–2518 (2015).

30. Falcaro, P. *et al.* Centimetre-scale micropore alignment in oriented polycrystalline metal–organic framework films via heteroepitaxial growth. *Nat. Mater.* **16**, 342–348 (2017).
31. Elnathan, R. *et al.* Biointerface design for vertical nanoprobes. *Nat. Rev. Mater.* 1–21 (2022) doi:10.1038/s41578-022-00464-7.
32. Tamames-Tabar, C. *et al.* Cytotoxicity of nanoscaled metal–organic frameworks. *J. Mater. Chem. B* **2**, 262–271 (2013).
33. Ettlinger, R. *et al.* Toxicity of metal–organic framework nanoparticles: from essential analyses to potential applications. *Chem. Soc. Rev.* **51**, 464–484 (2022).
34. Velic, A., Hasan, J., Li, Z. & Yarlagadda, P. K. D. V. Mechanics of Bacterial Interaction and Death on Nanopatterned Surfaces. *Biophys. J.* **120**, 217–231 (2021).
35. Rahimi, S. *et al.* Automated Prediction of Bacterial Exclusion Areas on SEM Images of Graphene–Polymer Composites. *Nanomaterials* **13**, 1605 (2023).
36. Chen, Y., Pandit, S., Rahimi, S. & Mijakovic, I. Graphene nanospikes exert bactericidal effect through mechanical damage and oxidative stress. *Carbon* **218**, 118740 (2024).
37. Kwon, O. *et al.* Computer-aided discovery of connected metal-organic frameworks. *Nat. Commun.* **10**, 3620 (2019).
38. Wang, X.-G., Xu, L., Li, M.-J. & Zhang, X.-Z. Construction of Flexible-on-Rigid Hybrid-Phase Metal–Organic Frameworks for Controllable Multi-Drug Delivery. *Angew. Chem. Int. Ed.* **59**, 18078–18086 (2020).
39. Liu, C., Wang, J., Wan, J. & Yu, C. MOF-on-MOF hybrids: Synthesis and applications. *Coord. Chem. Rev.* **432**, 213743 (2021).
40. Miyamoto, M., Kohmura, S., Iwatsuka, H., Oumi, Y. & Uemiya, S. In situ solvothermal growth of highly oriented Zr-based metal organic framework UiO-66 film with monocrystalline layer. *CrystEngComm* **17**, 3422–3425 (2015).
41. Lyu, D., Xu, W., Payong, J. E. L., Zhang, T. & Wang, Y. Low-dimensional assemblies of metal-organic framework particles and mutually coordinated anisotropy. *Nat. Commun.* **13**, 3980 (2022).
42. Wang, X., Cheng, H. & Zhang, X. Flexible-on-rigid heteroepitaxial metal-organic frameworks induced by template lattice change. *Nano Res.* (2021) doi:10.1007/s12274-021-4006-7.
43. Taddei, M. *et al.* Efficient microwave assisted synthesis of metal–organic framework UiO-66: optimization and scale up. *Dalton Trans.* **44**, 14019–14026 (2015).
44. Hu, Z. & Zhao, D. De facto methodologies toward the synthesis and scale-up production of UiO-66-type metal–organic frameworks and membrane materials. *Dalton Trans.* **44**, 19018–19040 (2015).
45. Wu, Y. *et al.* Optimized scalable synthesis and granulation of MIL-88B(Fe) for efficient arsenate removal. *J. Environ. Chem. Eng.* **10**, 108556 (2022).
46. Chakraborty, D., Yurdusen, A., Mouchaham, G., Nouar, F. & Serre, C. Large-Scale Production of Metal–Organic Frameworks. *Adv. Funct. Mater.* **34**, 2309089 (2024).
47. Products | ProfMOF - Metal Organic Frameworks. *https://profmof.com/products/* https://profmof.com/products/ (2024).
48. MIL-88B-Fe Powder MOF | Nanochemazone. https://www.nanochemazone.com/product/mil-88b-fe-mof/ (2024).
49. Yang, K. *et al.* Bacterial anti-adhesion surface design: Surface patterning, roughness and wettability: A review. *J. Mater. Sci. Technol.* **99**, 82–100 (2022).

50. Hu, Y., Dai, L., Liu, D. & Du, W. Rationally designing hydrophobic UiO-66 support for the enhanced enzymatic performance of immobilized lipase. *Green Chem.* **20**, 4500–4506 (2018).
51. Yang, C. Influence of Surface Roughness on Superhydrophobicity. *Phys. Rev. Lett.* **97**, (2006).
52. Zhu, R. *et al.* Fe-Based Metal Organic Frameworks (Fe-MOFs) for Bio-Related Applications. *Pharmaceutics* **15**, 1599 (2023).
53. Pandit, S. *et al.* Precontrolled Alignment of Graphite Nanoplatelets in Polymeric Composites Prevents Bacterial Attachment. *Small* **16**, 1904756 (2020).
54. Chen, Y. *et al.* Bioinspired nanoflakes with antifouling and mechano-bactericidal capacity. *Colloids Surf. B Biointerfaces* **224**, 113229 (2023).
55. Yi, Y. *et al.* Bioinspired nanopillar surface for switchable mechano-bactericidal and releasing actions. *J. Hazard. Mater.* **432**, 128685 (2022).
56. Kim, H.-K., Baek, H. W., Park, H.-H. & Cho, Y.-S. Reusable mechano-bactericidal surface with echinoid-shaped hierarchical micro/nano-structure. *Colloids Surf. B Biointerfaces* **234**, 113729 (2024).
57. Liu, Z. *et al.* Biocompatible mechano-bactericidal nanopatterned surfaces with salt-responsive bacterial release. *Acta Biomater.* **141**, 198–208 (2022).
58. Hawi, S. *et al.* Critical Review of Nanopillar-Based Mechanobactericidal Systems. *ACS Appl. Nano Mater.* **5**, 1–17 (2022).
59. Jenkins, J. *et al.* Antibacterial effects of nanopillar surfaces are mediated by cell impedance, penetration and induction of oxidative stress. *Nat. Commun.* **11**, 1626 (2020).
60. Zhao, S. *et al.* Programmed Death of Injured Pseudomonas aeruginosa on Mechano-Bactericidal Surfaces. *Nano Lett.* **22**, 1129–1137 (2022).
61. Elbourne, A. *et al.* Bacterial-nanostructure interactions: The role of cell elasticity and adhesion forces. *J. Colloid Interface Sci.* **546**, 192–210 (2019).

# Supplementary Information

**Mechano-bactericidal surfaces achieved by epitaxial growth of metal-organic frameworks**

Zhejian Cao, Santosh Pandit, Francoise M. Amombo Noa, Jian Zhang, Wengeng Gao, Shadi Rahimi, Lars Öhrström, and Ivan Mijakovic

E-mail: zhejian@chalmers.se, ohrstrom@chalmers.se, ivan.mijakovic@chalmers.se

# Contents

## Section 1: Natural and artificial mechano-bactericidal surfaces

The natural and artificial mechano-bactericidal (MB) surfaces are summarized in **Table S1**.

Table S1 Summary of natural and artificial mechano-bactericidal surfaces

| Surfaces | Preparation method | Surface features (height; tip/base diameter; pitch) | Wettability (water contact angle (CA)) | Tested species | Bactericidal activity | Refs |
|---|---|---|---|---|---|---|
| *Natural mechano-bactericidal surfaces* | | | | | | |
| Cicada wings (*P. claripennis*) | Natural | 200 nm; 60 nm / 100 nm; 170 nm | Hydrophobic, CA = 158° | *P. aeruginosa* | Lethal in 3 min | 1 |
| Dragonfly wing (*D. bipunctata*) | Natural | 240 nm; 50 nm; NA | Hydrophobic, CA = 153° | *P. aeruginosa, S. aureus, B. subtilis* | lethal effect to all 3 bacteria in 3 h | 2 |
| Gecko skin (*L. steindachneri*) | Natural | up to 4 µm; tip curvature in the range of 10–30 nm; sub-micron spacing | Hydrophobic, CA = 151-155° | *P. gingivalis* | Lethal effect up to 7 days | 3 |
| Gecko skin (*S. williamsi*) | Natural | 3 µm; 50 nm / 350-450 nm; 500 nm | Hydrophobic, CA = 134° | Co-culture of 7 gut bacteria | bactericidal function observed | 4 |
| Cicada wings (*M. intermedia*) | Natural | 241 nm; 156 nm; 165 | Hydrophobic, CA = 136° | *P. fluorescens* | Active at killing | 5 |
| Cicada wings (*C. aguila*) | Natural | 182 nm; 159 nm; 187 nm | Hydrophobic, CA = 113° | *P. fluorescens* | Active at killing | 5 |
| Cicada wings (*Magicicada ssp.*) | Natural | 84 nm; 167 nm; 252 nm | Hydrophilic, CA = 80° | *S. cerevisiae (Yeast)* | Cell wall rupture | 6 |
| Cicada wings (*Tibicen ssp.*) | Natural | 183 nm; 57 nm / 104 nm; 175 nm | Hydrophobic, CA = 132° | *S. cerevisiae (Yeast)* | Cell wall rupture | 6 |
| Dragonfly wings (*Progomphus. ssp.*) | Natural | 241 nm; 53 nm; 123 nm | Hydrophobic, CA = 119° | *S. cerevisiae (Yeast)* | Cell wall rupture | 6 |
| *Amorpha fruticosa* leaf | Natural | ~1 µm; randomly arranged | Hydrophobic, CA = 154° | *E. coli* | Cell rupture | 7 |
| *Artificial mechano-bactericidal surfaces (top-down approach: TD; bottom-up approach: BU)* | | | | | | |
| Black silicon (TD) | Reactive-ion etching (RIE) | 280 nm; 62 nm; 62 nm | Hydrophilic, CA = 8° | *P. aeruginosa, S. aureus* | 89% for *P. aeruginosa*; 85% for *S. aureus* in 24 h | 8 |
| Black silicon (TD) | RIE | 500 nm; 20-80 nm; NA | Hydrophilic, CA = 80° | *P. aeruginosa, S. aureus, B. subtilis* | lethal effect to all 3 bacteria in 3 h | 2 |
| Black silicon (TD) | RIE | NA; 150-200 nm; 100-250 nm | NA | *E. coli, S. aureus, B. cereus* | over 99% in 24h for all 3 bacteria | 9 |
| Black silicon (TD) | RIE | 4 µm; 10-20 nm / 220 nm; NA | Hydrophobic, CA = 154° | *E. coli, S. aureus* | 83% for *E. coli* and 86% for *S. aureus* in 3 h | 10 |
| Graphene nanosheets (BU) | Exfoliation + filtration | horizontal length 80 nm; width 5 nm; NA | NA | *P. aeruginosa, S. aureus* | 71% for *P. aeruginosa* and 77% for *S. aureus* | 11 |
| Graphene nanosheets (BU) | Chemical vapor deposition (CVD) | 60-100 nm; width less than 5 nm; NA | Hydrophobic, CA = 117° | *E. coli, S. epidermidis* | over 85% for *E. coli* and over 90% for *S. epidermidis* in 4 h | 12 |
| Carbon nanotubes (BU) | CVD | 1 µm; 50 nm; less than 10 nm | Hydrophobic, WCA = 149° | *P. aeruginosa, S. aureus* | 99.3% for *P. aeruginosa* and 84.9% for *S. aureus* | 13 |
| Titania (TD) | Hydrothermal etching | 3 µm; 100 nm; NA | NA | *P. aeruginosa, E. coli, S. aureus* | over 40% for *P. aeruginosa*, over 80% for *E. coli*; 5% for *S. aureus* | 14 |
| Titania (TD) | Hydrothermal etching | NA; 40 nm; NA | Hydrophilic, CA = 73° | *P. aeruginosa, S. aureus* | 47% for *P. aeruginosa* and 20% for *S. aureus* | 15 |
| Titania (TD) | RIE | 1 µm; 80 nm; NA | Hydrophilic, CA = 10° | *E. coli, P. aeruginosa, M. smegmatis, S. aureus* | 95% for *E. coli*, 98% for *P. aeruginosa*, 92% for *M. smegmatis*, and 22% for *S. aureus* in 4h | 16 |
| Titania (TD) | Hydrothermal etching | 4 µm; 50 nm / 100 nm; 3-5 µm | Hydrophilic, CA = 0° | *S. epidermidis* | 47% killing for *S. epidermidis* | 17 |
| Titania (BU) | Anodization | 2 µm; 10-30 nm; 2 µm | Hydrophilic, CA = 0° | *S. aureus* | 10-fold reduction | 18 |
| Gold (BU) | Anodization deposition | 100 nm; 50 nm; NA | NA | *S. aureus* | 99% for *S. aureus* | 19 |
| Ormostamp polymer (TD) | Templates by anodization and deep RIE | 150 nm - 400 nm; 80nm; 130 - 300 nm | NA | *S. aureus* | over 99% | 20 |
| PMMA (TD) | Stamping with molds | 210-300 nm; 70-215 nm; 100-380 nm | NA | *E. coli* | over 50% | 21 |
| **MOF-on-MOF (BU)** | **Solvothermal synthesis** | **300 nm; ~ 5 nm / 200 nm; ~ 500 nm** | **Hydrophobic, CA = 127°** | ***E. coli, S. epidermidis, S. aureus*** | **83% for *E. coli*** | **This work** |

## Section 2: Antibacterial metal-organic frameworks by metal ion release

The antibacterial metal-organic frameworks (MOFs) by metal ion release are summarized in **Table S2**.

Table S2 Summary of antibacterial MOFs by metal ion release

| MOFs | Antibacterial mechanism proposed | Tested species | Bactericidal activity | Refs |
|---|---|---|---|---|
| $Ag_2$(O-IPA)($H_2O$)·($H_3O$) | Ag ion release | E. coli, S. aureus | E. coli: MIC = 5–10 ppm; S. aureus: MIC = 10–15 ppm | 22 |
| $Ag_5$(PYDC)2(OH) | Ag ion release | E. coli, S. aureus | E. coli: MIC = 10–15 ppm; S. aureus: MIC = 15–20 ppm | 22 |
| $Ag_3$(3-phosphonobenzoate) | Ag ion release | P. aeruginosa, E. coli, S. aureus | P. aeruginosa: MBC = 20–30 µM; E. coli: MBC = 50 µM; S. aureus: MBC = 50–75 µM | 23 |
| [(AgL)$NO_3$]·$2H_2O$ | Ag ion release | E. coli, S. aureus | E. coli: MIC = 300 µM; S. aureus: MIC = 297 µM | 24 |
| [(AgL)$CF_3SO_3$]·$2H_2O$ | Ag ion release | E. coli, S. aureus | E. coli: MIC = 300 µM; S. aureus: MIC = 307 µM | 24 |
| [(AgL)$ClO_4$]·$2H_2O$ | Ag ion release | E. coli, S. aureus | E. coli: MIC = 308 µM; S. aureus: MIC = 293 µM | 24 |
| [$Ag_2$(µ-PTA)$_2$(µ-suc)]$_n$·$2nH_2O$, [$Ag_2$(µ-PTA)$_2$(µ$_4$-adip)]$_n$·$2nH_2O$, [$Ag_2$(µ$_4$-PTA)(µ$_4$-mal)]$_n$ | Ag ion release | P. aeruginosa, E. coli, S. aureus, C. albicans | P. aeruginosa: MIC = 6-20 µM; E. coli: MIC = 6-7 µM; S. aureus: MIC = 6-40 µM; C. albicans: MIC = 6-40 µM | 25 |
| [$Ag_7$(bte)$_4$($H_2O$) ($HP_2W^{VI}_{16}W^V_2O_{62}$)]·$2H_2O$, [$Ag_7$(btp)$_5$($HP_2W^{VI}_{16}W^V_2O_{62}$)]·$H_2O$, [$Ag_4$(btb)$_{3.5}$($P_2W_{18}O_{62}$)]($H_2$btb)·$2H_2O$ | Ag ion release | E. coli, S. aureus | Inhibition zone observed | 26 |
| CuBTC (HKUST-1) | Cu ion release | E. coli, S. aureus | Inhibition zone observed | 27 |
| CuBTC (HKUST-1) | Cu ion release | E. coli | Inhibition zone observed | 28 |
| CuBTC (HKUST-1) | Cu ion release | S. cerevisiae, G. candidum | S. cerevisiae: inhibition of growth in 24h, G. candidum: over 99% reduction in 24h | 29 |
| Cu-SURMOF 2 | Cu ion release | Cobetia marina | over 99% bacteria killed in 2h | 30 |
| n[Cu(AIP)$_2$(PIY)($H_2O$)$_2$]·$4H_2O$ | Cu ion release | P. aeruginosa, E. coli, Candida spp., S. aureus, Klebsiella sp. | Inhibition zone observed | 31 |
| MPN-DMOF* | Cu ion release | P. aeruginosa, B. vietnamensis, E. coli, and S. aureus | 1.6, 1.5, 0.9, and 1.1-log CFU reduction in the mature biofilms | 32 |
| ZnBDC (MOF-5) | Zn ion release | P. aeruginosa | P. aeruginosa: MIC = 25 µM | 33 |
| [Zn(bipy)(OH)$_2$)$_4^{2+}$]$_{1.5}$[$ClO_4^-$]$_3$·(bipy)$_3$($H_2O$) | Zn ion release | E. coli, S. epidermidis | E. coli: MIC = 5.3 ppm; S. epidermidis: MIC = 3.8 ppm | 34 |
| [$Zn_{1.5}$($CH_3CO_2$)$_2$(bipy)$^{2+}$][$ClO_4^-$]·$H_2O$ | Zn ion release | E. coli, S. epidermidis | E. coli: MIC = 6.1 ppm; S. epidermidis: MIC = 4.6 ppm | 34 |
| Zn(Im)2 (ZIF-4) | Zn ion release | E. coli, S. aureus | Inhibition zone observed for S. aureus | 35 |
| Zn(bIm)2 (ZIF-7)* | Zn ion release | E. coli, S. aureus | Inhibition zone observed for S. aureus | 35 |
| Zn(MeIm)2 (ZIF-8) | Zn ion release | E. coli, S. aureus | Inhibition zone observed | 35 |
| BioMIL-5 | Zn ion release | S. aureus, S. epidermidis | S. aureus: MIC =1.7 mg/mL; S. epidermidis: MIC =1.7 mg/mL | 36 |
| Co-TDM | Co ion release | E. coli | MBC=10-15 ppm | 37 |
| ZIF-67 | Co ion release | E. coli; P. putida; S. cerevisiae | Inhibition zone observed for both E. coli and P. putida | 38 |
| Co-SIM-1 | Co ion release | E. coli; P. putida; S. cerevisiae | Inhibition zone observed for both E. coli and P. putida | 38 |

*Some studies discussed the synergistic antimicrobial mechanisms of physical damage and other means, including positive charge[39]/cation release[32]. However, the quantitative contribution from the mechano-bactericidal part remains elusive. Furthermore, rational control of the surface features was not covered compared to the typical MB surfaces, such as tip size less than 200 nm, as summarized in **Table S1**.

## Section 3: Spacing analysis of the *in-situ* MIL-88B-on-UiO-66 (MoU) surfaces

The spacing/pitch distance of the nanostructures is one of the critical surface features for MB surfaces. The spacing of the vertical MIL-88B nanopillars is controlled by the distance between the center of UiO-66 cores, as the MIL-88B grew epitaxially from the UiO-66. The spacing between the adjacent MIL-88B nanopillars and UiO-66 cores in the *in-situ* surfaces is analyzed using ImageJ software. The results show that the spacing of MIL-88B nanopillars and UiO-66 cores are approximately 500 nm (527 ± 120 nm for MIL88-B and 542 ± 174 nm for UiO-66), as shown in **Table S3** and **S4**, **Figure S1** and **S2**.

Table S3 Spacing between vertical MIL-88B nanopillars

| Number | Length (nm) | Number | Length (nm) |
|---|---|---|---|
| 1 | 462 | 16 | 632 |
| 2 | 415 | 17 | 597 |
| 3 | 533 | 18 | 425 |
| 4 | 577 | 19 | 456 |
| 5 | 496 | 20 | 545 |
| 6 | 543 | 21 | 720 |
| 7 | 447 | 22 | 555 |
| 8 | 405 | 23 | 715 |
| 9 | 597 | 24 | 495 |
| 10 | 273 | 25 | 234 |
| 11 | 562 | 26 | 523 |
| 12 | 544 | 27 | 623 |
| 13 | 526 | 28 | 797 |
| 14 | 610 | **Mean** | **527** |
| 15 | 441 | **Standard deviation** | **120** |

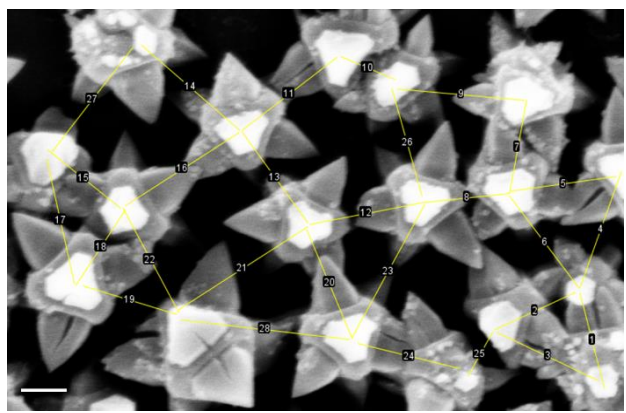

Figure S1 Spacing analysis of *in-situ* MoU surface, scale bar 200 nm

Table S4 Spacing between UiO-66 cores

| Number | Length (nm) | Number | Length (nm) |
|---|---|---|---|
| 1 | 517 | 21 | 902 |
| 2 | 547 | 22 | 1054 |
| 3 | 570 | 23 | 454 |
| 4 | 271 | 24 | 489 |
| 5 | 222 | 25 | 383 |
| 6 | 432 | 26 | 569 |
| 7 | 643 | 27 | 623 |
| 8 | 482 | 28 | 629 |
| 9 | 437 | 29 | 245 |
| 10 | 412 | 30 | 547 |
| 11 | 549 | 31 | 401 |
| 12 | 425 | 32 | 735 |
| 13 | 715 | 33 | 475 |
| 14 | 468 | 34 | 483 |
| 15 | 367 | 35 | 475 |
| 16 | 756 | 36 | 771 |
| 17 | 561 | 37 | 725 |
| 18 | 427 | 38 | 836 |
| 19 | 523 | **Mean** | **542** |
| 20 | 469 | **Standard deviation** | **174** |

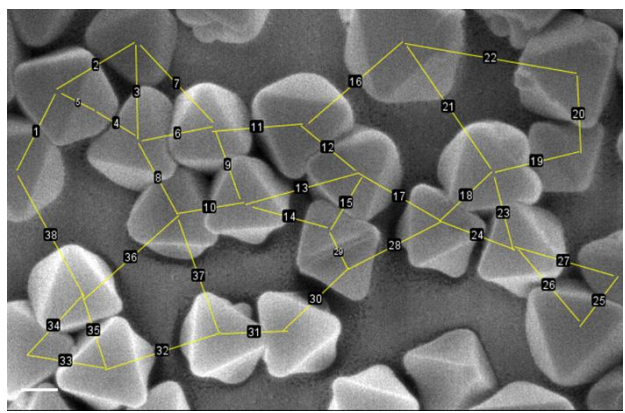

Figure S2 Spacing analysis of *in-situ* UiO-66 surface, scale bar 200 nm

## Section 4: Feasibility of modifying the MOF surface features

The surface features of the MB surfaces are essential to the MB performance.[40,41] One advantage of using MOF as the building block for MB surfaces is that there are various strategies to tune the geometry parameters of MOFs, such as adjusting the synthesis conditions.[42] More specifically, modulators play an important role in the nucleation and growth of MOF crystals.[43,44] By tuning the ratio of the modulator (in this study, acetic acid for UiO-66), we adjusted the size of the UiO-66 and achieved different diameters and lengths of the nanospikes in *in-situ* MoU surfaces as shown in **Figure S3a**. The pitch distance of the *in-situ* MoU is controlled by the *in-situ* UiO-66 cores, as discussed in **SI Section 3**. Many approaches to achieving a dense UiO-66 film were reported, such as adjusting the concentration of the modulator,[43] and repetition of the solvothermal treatments.[45] We have performed both methods to achieve a dense UiO-66 surface, as shown in **Figure S3bc**. The pitch distance and uncovered area could be reduced. However, using the repetition of solvothermal treatment (twice) might produce UiO-66 particles with different sizes (**Figure S3c**), resulting in a large standard deviation in the pitch distance. Our results verified the feasibility of rationally designing the surface features of the MOF MB surfaces, including pitch distance, diameter, and length of the nanospikes. However, precise control of the surface features requires systematic study with different synthesis conditions. Furthermore, an in-depth comprehensive study on the relationship between the geometry parameters of MOF surfaces and their bactericidal performance would be an important forward research direction, especially considering different bacteria possessing different geometries.

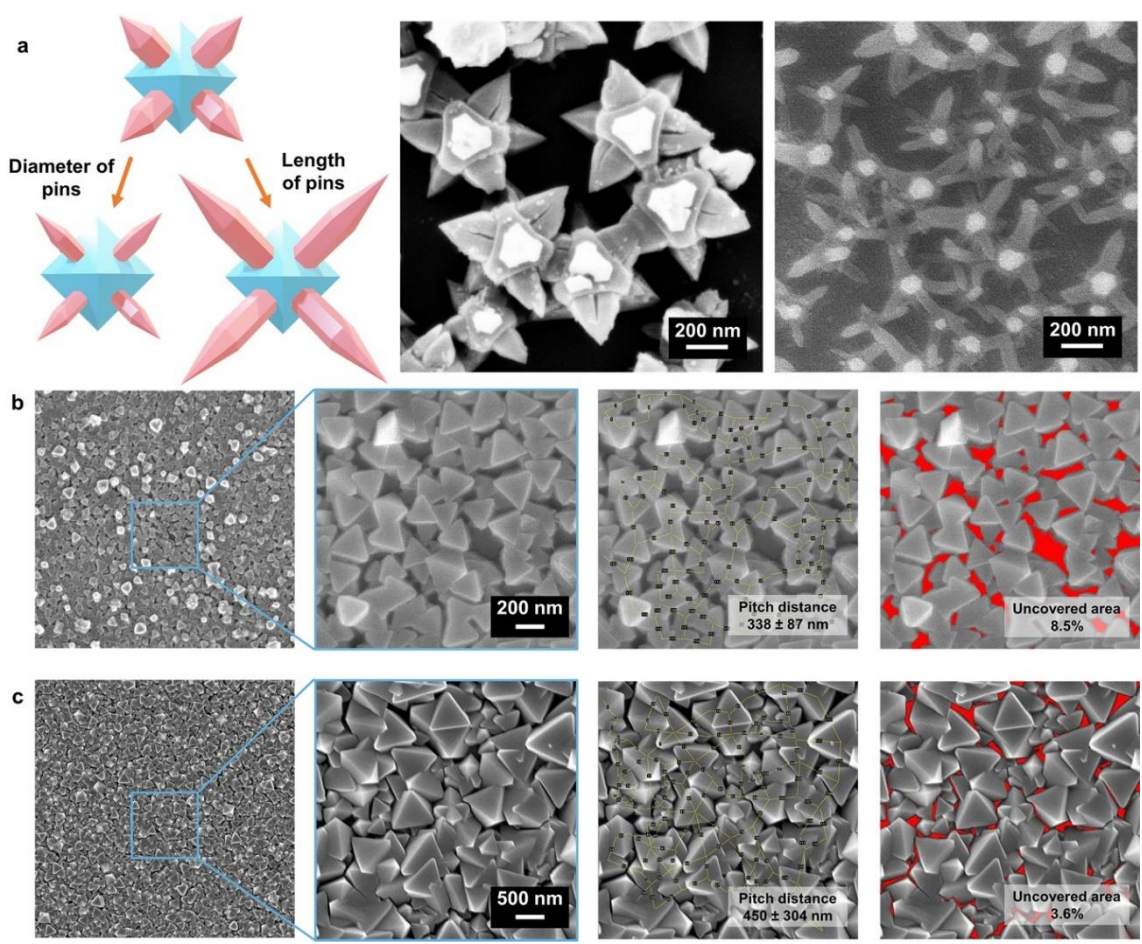

Figure S3 (a) Illustration and SEM images of the *in-situ* MoU surface with different diameters and lengths of the MIL-88B nanopillars; (b) dense *in-situ* UiO-66 surface achieved by tuning the modulator ratio; (c) dense *in-situ* UiO-66 surface achieved by two repetitions of solvothermal treatment

## Section 5: Dropcast MoU hybrids on different substrates

The melting point of typical medical plastics[46] and other materials[47] were summarized in **Table S5**. Both *in-situ* growth (120 °C) and *ex-situ* dropcasting (room temperature) methods work below the melting point of the common medical materials. Through an *ex-situ* method, MOF nanostructures can be loaded on various substrates. For instance, the MoU hybrids were loaded on different substrates by dropcasting as shown in **Figure S4**. Four substrates including glass, stainless steel, wood, and polyethylene terephthalate (PET) slice have been used as the substrates. Moreover, pilot-scale production of UiO-66(Zr) has been reported with over 93% yield,[48] and the production of MIL-88B(Fe) has been scaled up by replacing DMF with ethanol.[49] The low fabrication temperature of MB surface assembly and reported scale-up cases indicate promising potential for large-scale production of MoU MB surfaces on different substrates.

Table S5 Melting temperature of common medical materials

|  | Materials | Melting temperature (°C) |
|---|---|---|
| **Polymers** | High-density polyethylene (HDPE) | 134 |
|  | Polypropylene (PP) | 170 |
|  | Polyvinyl chloride (PVC) | 170 |
|  | Polyamides (Nylon 6) | 215 |
|  | Polyethylene terephthalate (PET) | 255 |
|  | polytetrafluorethylene (PTFE) | 327 |
| **Metals** | Gold | 1064 |
|  | Stainless steel (316) | 1375 |
|  | Titanium | 1668 |
| **Ceramics** | Glass | 1400 |
|  | Aluminum oxide | 2072 |
|  | Zirconium oxide | 2715 |

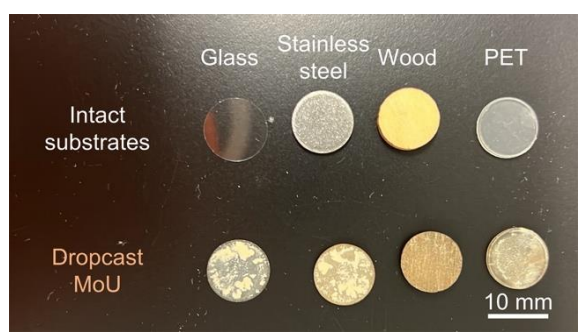

Figure S4 Dropcast MoU on different substrates. The first row is the intact substrates, and the second row is the samples with dropcast MoU. The substrates are from left to right: glass, stainless steel, wood, and PET slice

## Section 6: Simulated XRD patterns of UiO-66 and MIL-88B

The original structure Crystallographic Information Files (CIF) were obtained from Cambridge Crystallographic Data Centre "CCDC" with identifiers "RUBTAK04" and "YEDKOI" for UiO-66 and MIL-88B, respectively. Their structures and XRD patterns are illustrated in **Figure S5**. The XRD patterns were simulated in Mercury software.

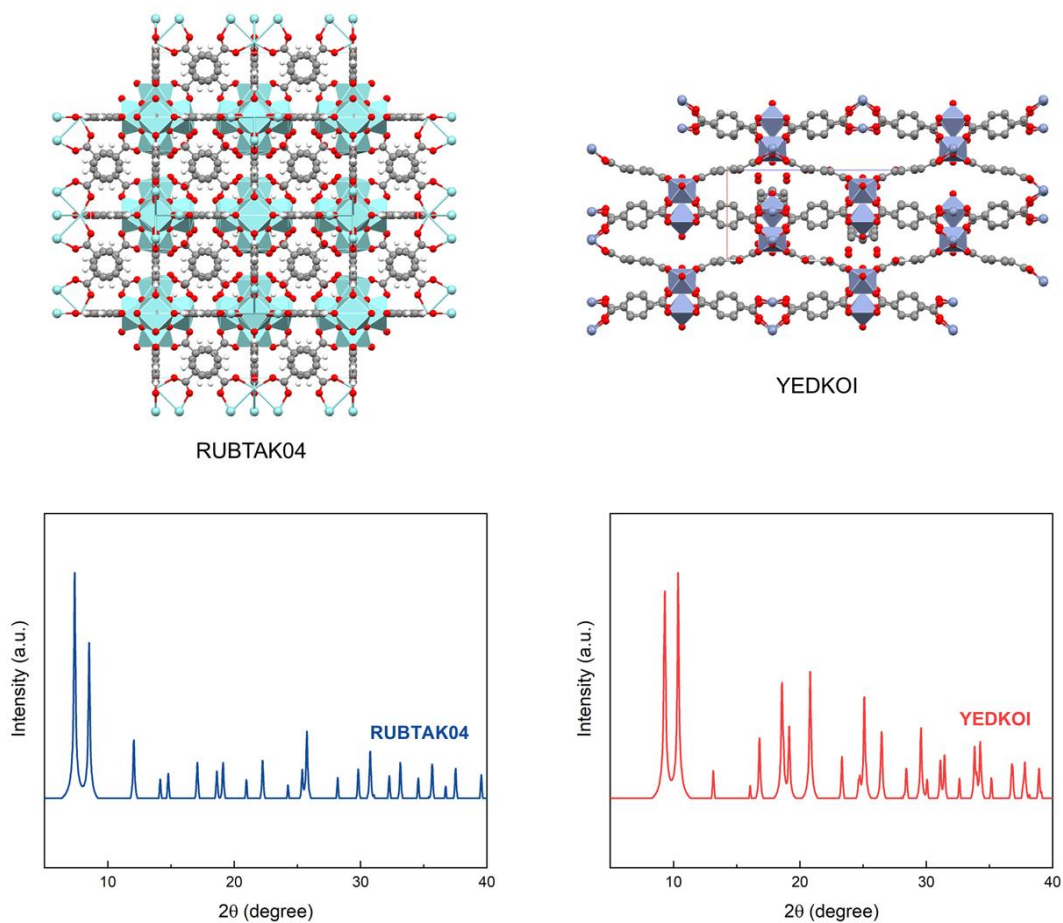

Figure S5 Structures and XRD patterns of RUBTAK04 (UiO-66) and YEDKOI (MIL-88B)

## Section 7: High-resolution XPS spectra for Fe *2p* and Zr *3d*

The high-resolution XPS spectra for Fe *2p* and Zr *3d* are shown in **Figure S6**. Fe *2p* peaks were observed in the MIL-88B, *in-situ* MoU, and dropcast MoU samples, as shown in **Figure S6a**. The main Fe *2p* peaks centered at 725.3 eV and 712.0 eV, corresponding to Fe $2p_{1/2}$ and Fe $2p_{3/2}$, respectively.[50] Two satellite (sat.) signals were observed near 731.1 eV and 717.5 eV. These Fe *2p* peaks confirmed the existence of $Fe^{3+}$.[51] Zr *3d* peaks were observed in the UiO-66, *in-situ* MoU, and dropcast MoU samples, as shown in **Figure S6b**. The main Zr *3d* peaks centered at 184.9 eV and 182.6 eV, corresponding to Zr $3d_{3/2}$ and Zr $3d_{5/2}$, respectively. These Zr *3d* peaks confirmed the existence of $Zr^{4+}$.[52]

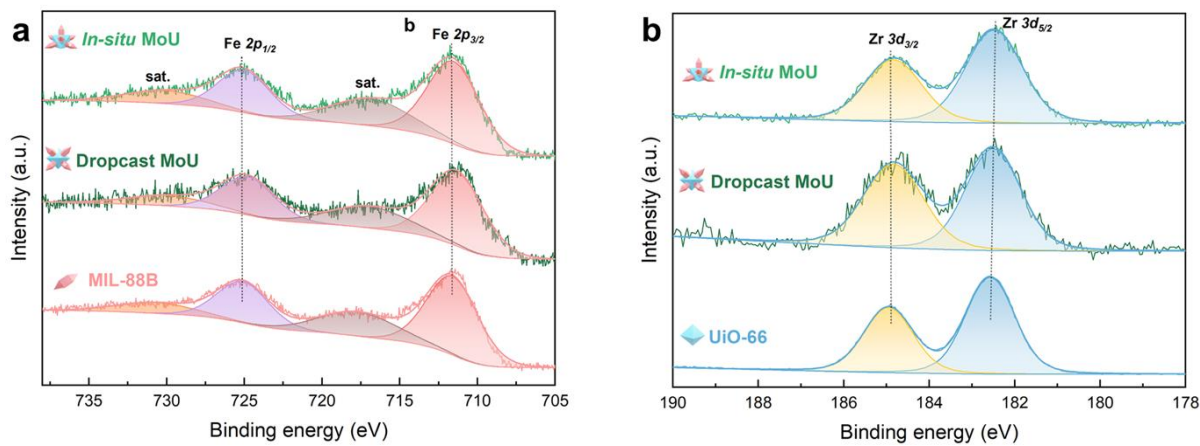

Figure S6 XPS spectra of (a) Fe *2p* and (b) Zr *3d*

## Section 8: Water contact angle measurement

The wettability of the obtained surfaces was analyzed by water contact angle (CA) measurement within 5s once the water contacted the surfaces. The hydrophobicity of the surface was enhanced when introducing MOFs, as shown in **Table S6** and **Figure S7**.

Table S6 Water contact angle of obtained surfaces

| Number | Si substrate | | | Glass substrate | | | | |
|---|---|---|---|---|---|---|---|---|
| | Si | *In-situ* UiO-66 | *In-situ* MoU | Glass | Dropcast UiO-66 | Dropcast MIL-88B | Dropcast MIL-88B + UiO-66 | Dropcast MoU |
| 1 | 56 | 99 | 126 | 60 | 129 | 124 | 124 | 125 |
| 2 | 55 | 102 | 129 | 61 | 121 | 125 | 124 | 122 |
| 3 | 54 | 105 | 125 | 63 | 121 | 123 | 126 | 126 |
| Mean | 55 | 102 | 127 | 61 | 124 | 124 | 125 | 124 |

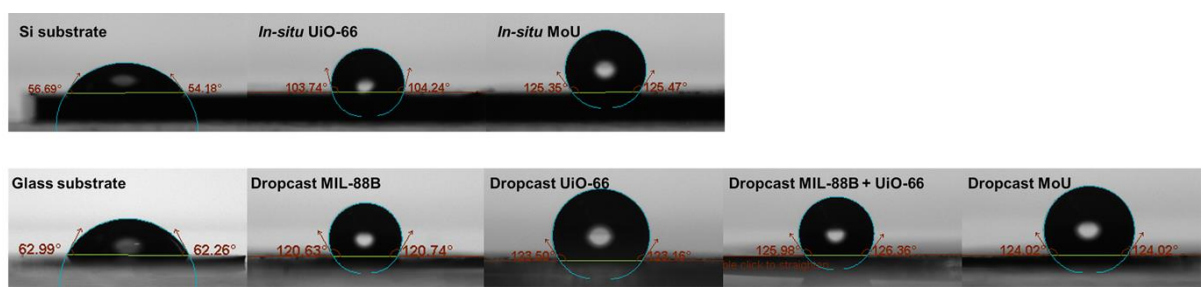

Figure S7 Images of water contact angle measurements for obtained surfaces

## Section 9: Zone of inhibition test

The zone of inhibition tests were carried out with *E. coli* and *S. aureus* in three biological replicates to preliminarily evaluate the potential chemical leaching and chemical antibacterial activities. Ampicillin was used at the same loading concentration with MoU as a demonstration of the zone of inhibition. As shown in **Table S7** and **Figure S8**, no inhibition zone was observed for obtained MOF surfaces.

Table S7 Zone of inhibition results for the tested MOF surfaces

| Time | ZOI with *E. coli* (mm) | | | | | ZOI with *S. aureus* (mm) | | | | |
|---|---|---|---|---|---|---|---|---|---|---|
| | Si | MoU | UiO-66 | MIL-88B | Ampicillin | Si | MoU | UiO-66 | MIL-88B | Ampicillin |
| 12 h | 0 | 0 | 0 | 0 | 26 | 0 | 0 | 0 | 0 | 45 |
| 24 h | 0 | 0 | 0 | 0 | 26 | 0 | 0 | 0 | 0 | 50 |
| 48 h | 0 | 0 | 0 | 0 | 26 | 0 | 0 | 0 | 0 | 51 |
| 72 h | 0 | 0 | 0 | 0 | 26 | 0 | 0 | 0 | 0 | 51 |

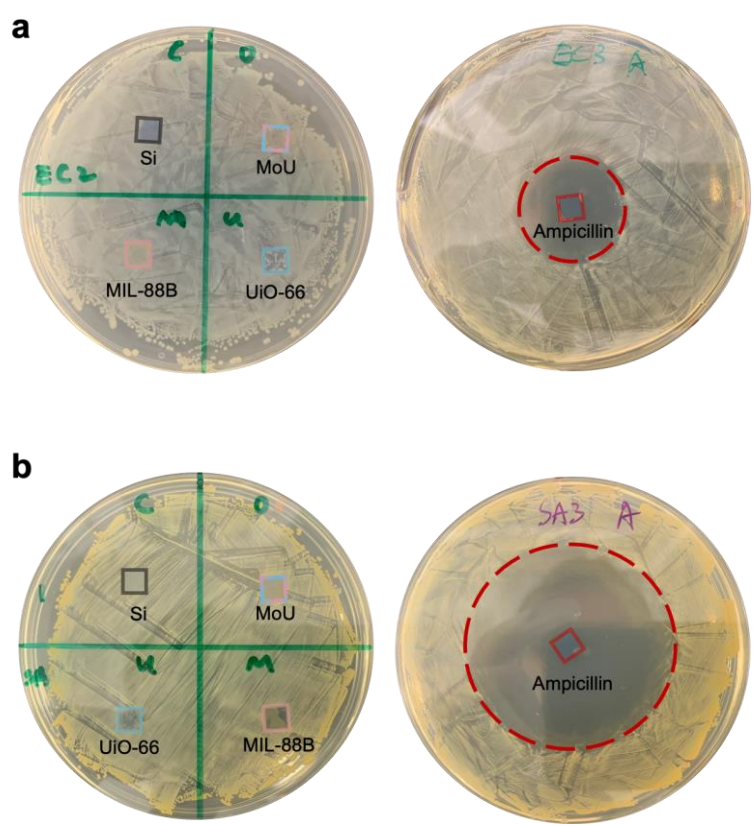

Figure S8 Images of zone of inhibition (ZOI) test with 24h growth of (a) *E. coli* and (b) *S. aureus;* the boundaries of tested samples were highlighted with square frames; ZOI was found in ampicillin samples, marked with dashed circles

## Section 10: Bactericidal efficiency for Gram-positive bacteria

The antibacterial performance of the MOF MB surface towards Gram-positive bacteria was evaluated by colony forming unit (CFU) counting method and live/dead staining, as shown in **Figure S9**. The surfaces containing MIL-88B presented a considerable reduction in the CFU counting. For example, dropcast MIL-88B demonstrated 97.7% and 99.9% bactericidal efficiency for *S. epidermidis* and *S. aureus*, respectively. This excellent bactericidal efficiency was not only owing to the mechano-bactericidal effect but also synergistically contributed by the $Fe^{3+}$ release, as the degradation of the MIL-88B was found in the SEM images when Gram-positive bacteria growing with liquid culture media, as shown in **Figure S10**. The degradation of MIL-88B could release large amounts of $Fe^{3+}$, and these released $Fe^{3+}$ could thereby participate in Fenton-like reactions and result in antibacterial effects.[53,54] Therefore, we observed high bactericidal efficiency of MoU (99% for *S. aureus*) and M+U surfaces for Gram-positive bacteria. Notably, mechano-bactericidal effects of the MoU surfaces towards Gram-positive bacteria were still observed in the SEM images as described in the main text. However, distinguishing the contribution of the mechano-bactericidal part and the Fenton-like reaction part will require further study, and *E. coli* was used as the model for the quantitative study of MB actions of the MOF surfaces in this work.

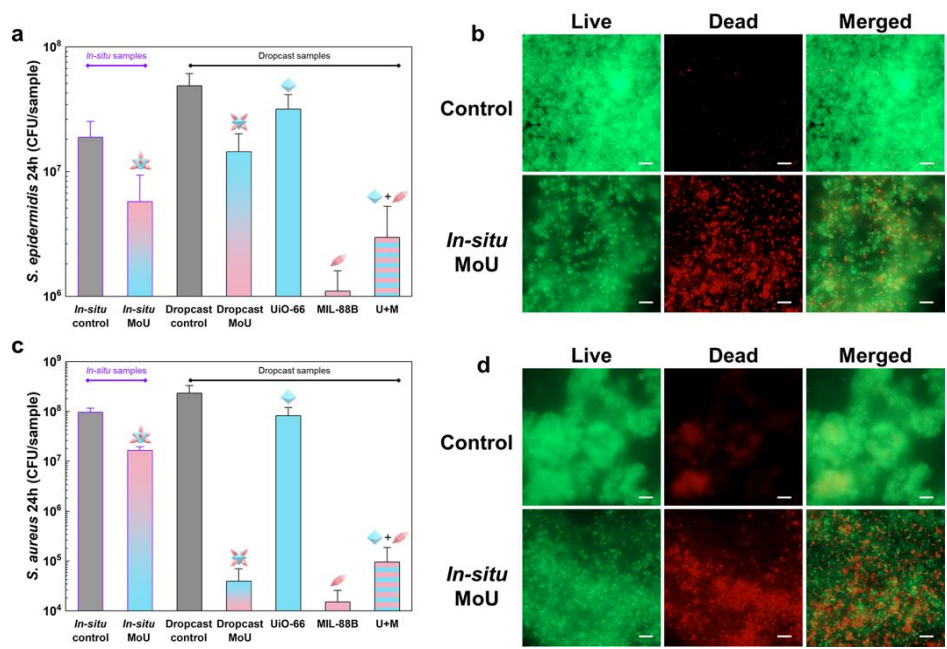

Figure S9 CFU results of attached bacteria after 24h growth: (a) *S. epidermidis* on *in-situ* MoU samples and dropcasting samples, including MoU, UiO-66, MIL-88B, and UiO-66 + MIL-88B (U+M); (c) *S. aureus* on *in-situ* MoU samples and dropcasting samples, including MoU, UiO-66, MIL-88B, and U+M, Data represent the mean ± standard deviation of three biological replicates. The live/dead fluorescent staining images of attached bacteria with 24h growth: (b) *S. epidermidis* on control and *in-situ* MoU surfaces; (d) *S. aureus* on control and *in-situ* MoU surfaces, green indicating live bacteria and red indicating dead bacteria, scale bar: 10 μm

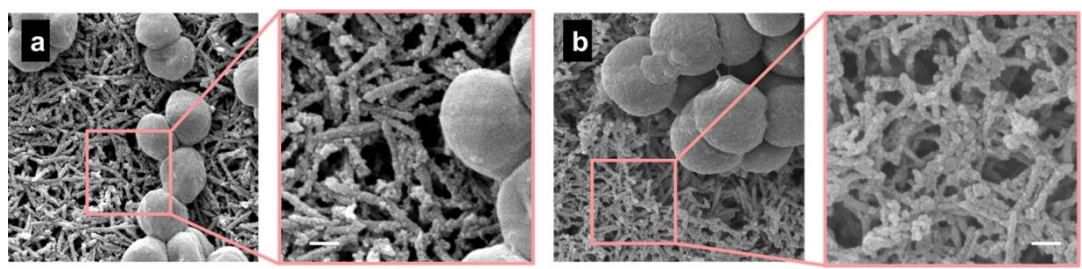

Figure S10 SEM images of dropcast MIL-88B surfaces after Gram-positive bacteria growth for 24h: (a) *S. epidermidis*, (b) *S. aureus*; degradation of MIL-88B was observed, scale bar: 200 nm

**Section 11: Effective bactericidal area analysis**

The effective bactericidal area was analyzed by the previously reported methods.[55,56] SEM images of *in-situ*/dropcast MoU surfaces with three different regions were used for the uncovered area analysis. Dropcast MoU surfaces possess less gap areas (15%) than the *in-situ* MoU surfaces (36%), with detailed information in **Table S8**. Furthermore, due to the random four-pin up orientation of the nanopillars in dropcast MoU, more MIL-88B nanopillars of the MoU could also induce mechanical stress on bacteria, which could contribute to an elevated bactericidal efficiency in dropcast MoU surfaces compared to the *in-situ* MoU surfaces.

Table S8 The percentage of gap areas on *in-situ* and dropcast MoU surfaces

| No. | *In-situ* | Dropcast |
|---|---|---|
| 1 | 33% | 14% |
| 2 | 39% | 20% |
| 3 | 37% | 12% |
| Mean | 36% | 15% |
| Stand. Div. | 2.49% | 3.40% |

## Section 12: Bactericidal efficiency of dropcast MoU surface for 72h

To investigate the long-term bactericidal performance of the MOF surfaces, plate counting of colony-forming unit (CFU) method was used to evaluate the attached bacteria on the dropcast MoU surfaces after 72h growth of *E. coli*. The LB culture medium was changed every 24 hours to provide fresh nutrients. As shown in **Figure S11**, the MoU surfaces demonstrated relatively higher bactericidal efficiency (51%) than UiO-66 (17%), MIL-88B (31%), and UiO-66 + MIL-88B (U+M) (9%). However, the bactericidal efficiency dropped from 83% (24h growth) to 51% (72h growth), which could be attributed to the coverage of the nanostructures by the debris. The debris including the dead bacteria could impede the contact of the planktonic bacteria to the nanostructures of the MB surfaces, leading to diminished bactericidal performance for certain MB surfaces.[57,58] Thus, a strategy to efficiently clean the MB surfaces would be required to provide long-term protection for our MOF MB surfaces.[59]

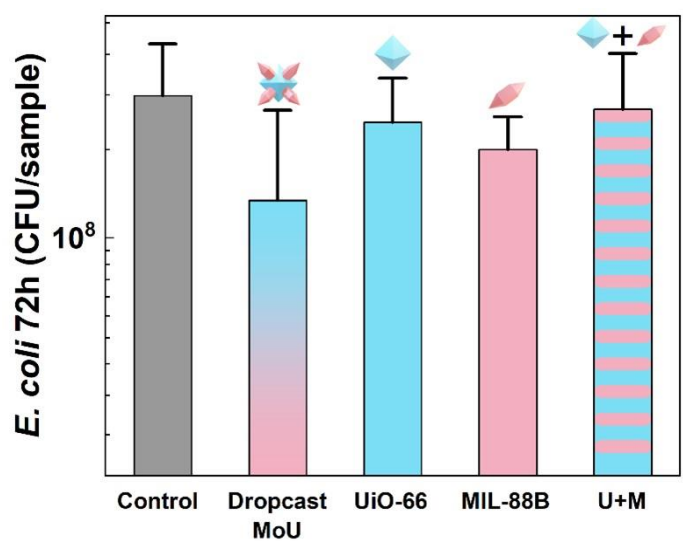

Figure S11 CFU results of attached *E. coli* with 72h growth on dropcasting samples, including MoU, UiO-66, MIL-88B, and UiO-66 + MIL-88B (U+M), Data represent the mean ± standard deviation of three biological replicates

**Section 13: Tilted SEM images of *in-situ* MOF MB surfaces**

To achieve better observation of the *in-situ* MOF MB surfaces as well as the interaction between the MOF structures and bacteria, tilted (45°) SEM images were acquired as shown in **Figure S12**. Tilted SEM images of the *in-situ* UiO-66 surfaces confirmed UiO-66 seamlessly grew on the substrate and provided a near horizontal triangle surface for the following MoU epitaxial growth. Tilted SEM images of *in-situ* MoU surfaces confirmed the one-pin up orientation of the MoU hybrid. Furthermore, with close-up views of *E. coli* and *S. aureus*, the MB actions were verified, as the *stretching*, *impaling,* and *apoptosis-like death induced by mechanical injury* were observed with a lateral view with torn and deflated bacterial envelopes, as highlighted in the yellow dashed circles.

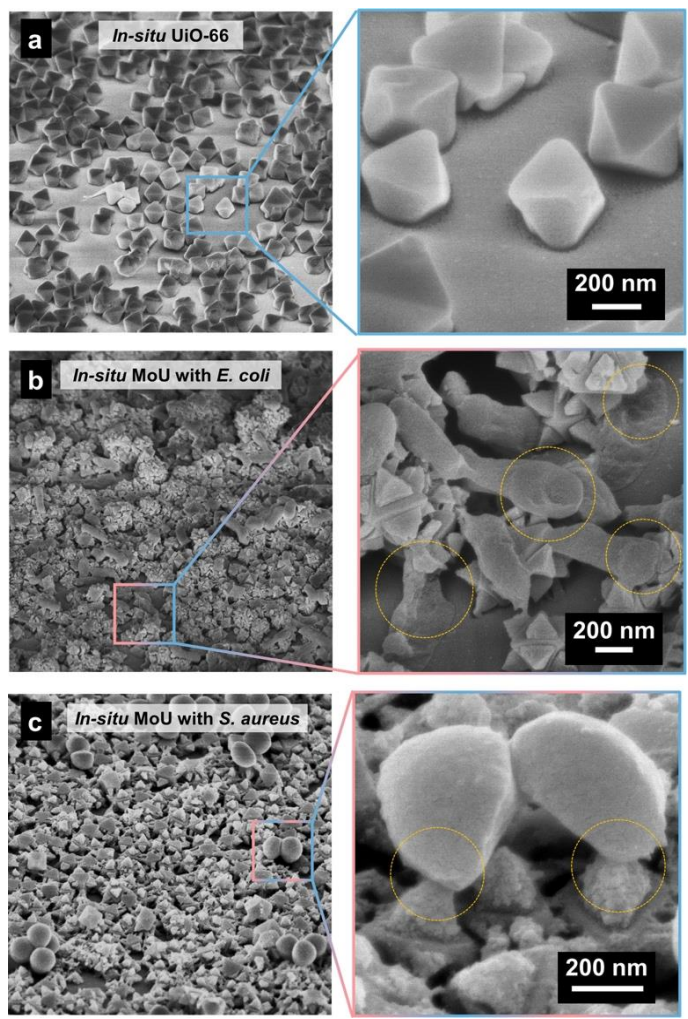

Figure S12 Tilted (45°) SEM images of (a) *in-situ* UiO-66 surfaces, *in-situ* MoU surfaces with attached (b) *E. coli* and (c) *S. aureus* after 24h growth; mechano-bactericidal actions were observed and highlighted in the yellow dashed circles

S14

## Section 14: SEM images of attached bacteria on MOF surfaces

SEM is a great tool to reveal the interaction of bacteria and MOF surfaces. MB actions were found not only for *E. coli* and *S. epidermidis* but also for MDR *S. aureus* in both *in-situ* MoU and dropcast MoU (**Figure S13ab**), where direct impaling and mechanical stress were observed. However, the MB actions were rarely found on the sole UiO-66 and MIL-88B surfaces due to missing critical geometry features of MB surfaces, such as sharp tips and vertical orientation, as shown in **Figure S13 (c) – (f)**. Degradation of the MIL-88B was found in the Gram-positive bacteria samples as described in Section 10.

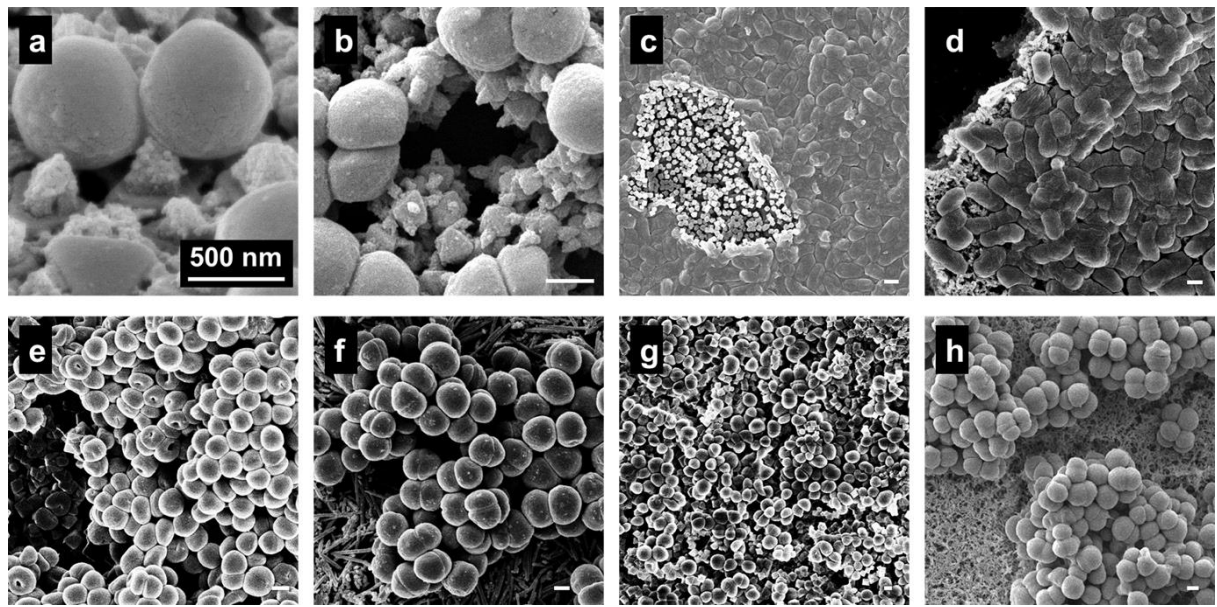

Figure S13 SEM images of attached bacteria on different MOF surfaces after 24h growth. *S. aureus* on (a) *in-situ* and (b) dropcast MoU surfaces, mechanically inducing deformation presented on *S. aureus,* and degradation of the MIL-88B part of the MoU was observed. Attached bacteria on dropcast UiO-66 and MIL-88B with: *E. coli* (c) and (d); *S. epidermidis* (e) and (f); *S. aureus* (g) and (h). Mechano-bactericidal actions were rarely observed on dropcast UiO-66 or MIL-88B surfaces. Scale bar: 500 nm

## Section 15: Stability of the MoU hybrid and MoU surfaces

Good stability of MoU hybrid and MoU surfaces is required to provide reliable protection from bacterial attachment. For the MoU hybrid, the seamless interfaces have been verified in previously reported epitaxial MOF growth work,[60] confirming a stable connection between the UiO-66 cores and MIL-88B satellites. Our synthesized MoU hybrid presented excellent structural stability, revealed by the preserved intact structure after ultrasonication treatment and immersion in water, culture media, and ethanol, as shown in **Figure S14a**. We also verified the stability of the MoU surfaces by SEM images after different treatments. Ultrasonication was applied to remove the dangling MOFs and debris after the synthesis of *in-situ* MoU surfaces. MoU hybrids maintained full coverage on the silicon surface after sonification and bio-tests, indicating strong adhesion between the MOFs and substrates (**Figure S14b**). For the *ex-situ* dropcast MoU surface, the same fixation protocol (80 °C, 3h) was applied according to our previous work.[12,56,61] The obtained *ex-situ* MoU dropcast surfaces maintained full coverage over the glass slice and were not broken nor washed away after bactericidal evaluation and series dehydration (40%, 50%, 60%, 70%, 80%, 90%, and 100% ethanol), where several times of pipetting were involved (**Figure S14c**). The intact structure of the MoU hybrids and MoU surfaces confirmed good stability and adhesion required for the MB action study in this work. However, for applications require strong adhesion, adhesive layers, such as polydopamine, hydrogel, could be introduced as reported in many works.[62,63]

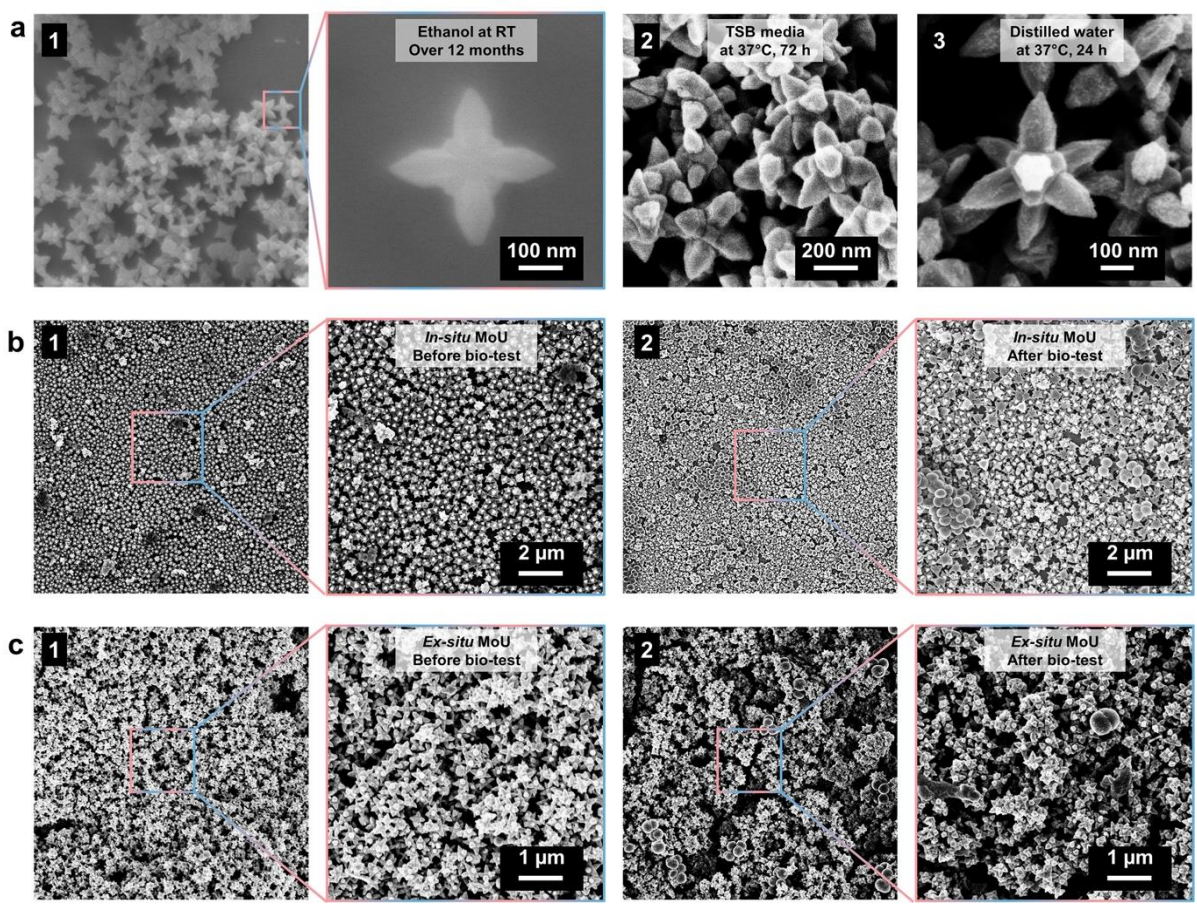

Figure S14 SEM images of MoU hybrids and surfaces. (a) MoU hybrids maintained their intact structures after immersion in solvents under different conditions: (1) ethanol at room temperature (RT, around 20 °C) for 12 months, (2) TSB media at 37 °C for 72 h, (3) distilled water at 37 °C for 24 h. (b) *In-situ* MoU surfaces maintained full surface coverage: (1) before bio-tests, (2) after bio-tests (24h, 37 °C, with *S. epidermidis*) with several times of pipetting. (c) (b) *Ex-situ* dropcast MoU surfaces maintained full surface coverage: (1) before bio-tests, (2) after bio-tests (72h, 37 °C, with *S. epidermidis*) with several times of pipetting

## Section 16: Stress analysis simulation

The *in-situ* MoU surfaces were used for the stress analysis as the geometry features can be obtained through SEM and TEM images. The model of the *in-situ* MoU surface was first simplified as shown in **Figure S15ab**. *E.coli* and *S. aureus* were selected as the bacteria models as they were tested in this work and with many reported parameters, such as adhesion force.[64–66] The simulation was performed using the software Ansys 2024 R1 as the environment of the finite element analysis. The size information and other parameters used in the simulation are summarized in **Table S9**. Notably, the chemical interactions between the MOF particles and the bacterial lipids were not taken into consideration in this simulation. However, the MOF surfaces have been proven to be hydrophobic in Section 6, and the hydrophobicity of the nanostructures has been reported to be beneficial for penetration into bacteria.[56,61]

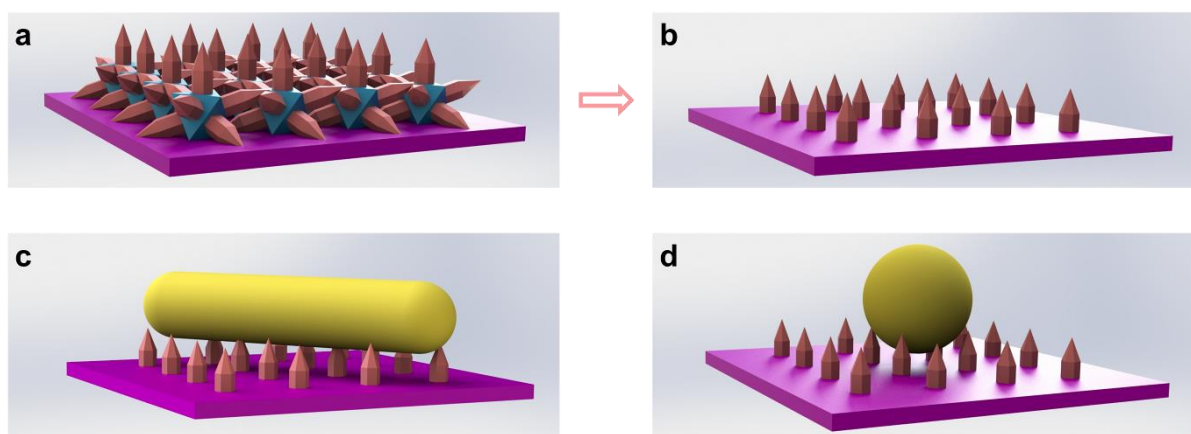

Figure S15 Structure modeling of the (a) *in-situ* MoU surface, (b) simplified *in-situ* MoU surface; (c) *E. coli* and (d) *S. aureus* on simplified *in-situ* MoU surface

Table S9 Parameters for numerical computation of stress distribution for bacteria on MOF MB surfaces

| Object | Parameters |
|---|---|
| *E. coli* | Cylinder diameter: 1 μm |
| | Cylinder length: 2 μm |
| | Hemisphere diameter: 500 nm |
| | Density: 1105 kg m$^{-3}$[67] |
| | Young's modulus: 5 MPa[68] |
| | Adhesion force: 10 nN[69] |
| *S. aureus* | Sphere diameter: 800 nm |
| | Density: 1105 kg m$^{-3}$ |
| | Young's modulus: 1.5 MPa[68] |
| | Adhesion force: 10 nN[66] |
| **MOF surface (MIL-88B nanopillar)** | Side length of hexagon: 97 nm |
| | Height of the Hexagonal prism: 131 nm |
| | Length of the tip: 172 nm |
| | Pitch distance: 527 nm |
| | Young's modulus: 7.7 GPa[70] |

### Section 17: Fabrication of MOF MB surfaces

All the starting chemicals, including $ZrCl_4$, $Fe(NO_3)_3 \cdot 9H_2O$, Benzene-1,4-dicarboxylic acid (BDC), acetic acid (AcOH), acetonitrile ($CH_3CN$), and *N,N*-Dimethylformamide (DMF) were obtained commercially (Sigma Aldrich) and used as received without further purification.

UiO-66, MIL-88B, and MIL-88B on UiO-66 (MoU) were obtained by modifying previous work, as listed in detail below.

<u>UiO-66(Zr)</u>: UiO-66 was synthesized based on a reported method.[71] $ZrCl_4$ solution (DMF, 36 mM) and BDC solution (DMF, 36 mM) were mixed, and then AcOH was added to the solution for a 10 min stirring mixing at room temperature. The molar ratio of $ZrCl_4$ : BDC : DMF: AcOH is 1:1:720:580. The obtained mixture was then transferred to a Teflon-lined autoclave. Silicon chips (6 mm * 6mm) were loaded to the bottom of the autoclave to obtain the *in-situ* UiO-66 surfaces[45]. The autoclave was held at 120 °C for 24h. After cooling to room temperature, the obtained powder was filtered and subsequently washed by DMF 3 times and ethanol 3 times through redispersion. The *in-situ* UiO-66 samples were washed by DMF 3 times and ethanol 3 times through pipette rinsing, followed by sonification at 80% power to remove the dangling particles. Obtained UiO-66 powder and *in-situ* UiO-66 were dried in a static vacuum oven at 60 °C overnight.

<u>MIL-88B(Fe)</u>: MIL-88B was synthesized based on a reported method.[72] $Fe(NO_3)_3 \cdot 9H_2O$ solution (DMF, 200 mM) and BDC solution (DMF, 100 mM) were mixed, and then $CH_3CN$ was added to the solution for a 10 min stirring mixing at room temperature. The molar ratio of $Fe(NO_3)_3 \cdot 9H_2O$ : BDC : DMF: $CH_3CN$ is 1:1:200:290. The obtained mixture was then transferred to Pyrex tubes and held at 90 °C for 5h. After cooling to room temperature, the obtained powder was filtered and subsequently washed by DMF 3 times and ethanol 3 times through redispersion. Obtained MIL-88B powder was dried in a static vacuum oven at 60 °C overnight.

<u>MIL-88B on UiO-66 (MoU)</u>: MoU was synthesized based on reported methods.[60,72] $Fe(NO_3)_3 \cdot 9H_2O$ solution (DMF, 200 mM, 3 mL) and BDC solution (DMF, 100 mM, 6mL) were mixed, and then $CH_3CN$ (9 mL) was added to the solution for a 10 min stirring mixing at room temperature. Thereafter, 43 mg UiO-66 (in 2 mL DMF) was added to the mixture, followed by 5 min sonification. The obtained mixture was then transferred to Pyrex tubes where an *in-situ* UiO-66 chip was loaded to the bottom of the tubes and then held at 90 °C for 5h. After cooling to room temperature, the obtained powder was filtered and subsequently washed by DMF 3 times and ethanol 3 times through redispersion. The *in-situ* MoU samples were washed by DMF 3 times and ethanol 3 times through pipette rinsing, followed by sonification at 80% power to remove the dangling particles. Obtained MoU powder and in-situ MoU were dried in a static vacuum oven at 60 °C overnight.

<u>Dropcast MOF MB surfaces</u>: MoU, UiO-66, and MIL-88B were dispersed in ethanol at a concentration of 5 mg mL$^{-1}$. UiO-66 and MIL-88B solution were mixed at the volume ratio of 1:1, denoted as MIL-88B plus UiO-66 (M+U). After sonification, 50 μL MOF solution was dropcast on the round glass slide (diameter 1 cm). After drying at room temperature, the samples were loaded in an 80 °C oven for 3 h to fix the MOFs to the surfaces. All the samples were dried in a static vacuum oven at 60 °C overnight before tests.

**Reference for supplementary information**

1. Ivanova, E. P. *et al.* Natural Bactericidal Surfaces: Mechanical Rupture of Pseudomonas aeruginosa Cells by Cicada Wings. *Small* **8**, 2489–2494 (2012).
2. Ivanova, E. P. *et al.* Bactericidal activity of black silicon. *Nat. Commun.* **4**, 2838 (2013).
3. Watson, G. S. *et al.* A gecko skin micro/nano structure – A low adhesion, superhydrophobic, anti-wetting, self-cleaning, biocompatible, antibacterial surface. *Acta Biomater.* **21**, 109–122 (2015).
4. Li, X. *et al.* The nanotipped hairs of gecko skin and biotemplated replicas impair and/or kill pathogenic bacteria with high efficiency. *Nanoscale* **8**, 18860–18869 (2016).
5. Kelleher, S. M. *et al.* Cicada Wing Surface Topography: An Investigation into the Bactericidal Properties of Nanostructural Features. *ACS Appl. Mater. Interfaces* **8**, 14966–14974 (2016).
6. Nowlin, K., Boseman, A., Covell, A. & LaJeunesse, D. Adhesion-dependent rupturing of Saccharomyces cerevisiae on biological antimicrobial nanostructured surfaces. *J. R. Soc. Interface* **12**, 20140999 (2015).
7. Chen, Y. *et al.* Bioinspired nanoflakes with antifouling and mechano-bactericidal capacity. *Colloids Surf. B Biointerfaces* **224**, 113229 (2023).
8. Linklater, D. P., Nguyen, H. K. D., Bhadra, C. M., Juodkazis, S. & Ivanova, E. P. Influence of nanoscale topology on bactericidal efficiency of black silicon surfaces. *Nanotechnology* **28**, 245301 (2017).
9. Vassallo, E. *et al.* Bactericidal performance of nanostructured surfaces by fluorocarbon plasma. *Mater. Sci. Eng. C* **80**, 117–121 (2017).
10. Hasan, J., Raj, S., Yadav, L. & Chatterjee, K. Engineering a nanostructured "super surface" with superhydrophobic and superkilling properties. *RSC Adv.* **5**, 44953–44959 (2015).
11. Pham, V. T. H. *et al.* "Race for the Surface": Eukaryotic Cells Can Win. *ACS Appl. Mater. Interfaces* **8**, 22025–22031 (2016).
12. Pandit, S. *et al.* Vertically Aligned Graphene Coating is Bactericidal and Prevents the Formation of Bacterial Biofilms. *Adv. Mater. Interfaces* **5**, 1701331 (2018).
13. Linklater, D. P. *et al.* High Aspect Ratio Nanostructures Kill Bacteria via Storage and Release of Mechanical Energy. *ACS Nano* **12**, 6657–6667 (2018).
14. Diu, T. *et al.* Cicada-inspired cell-instructive nanopatterned arrays. *Sci. Rep.* **4**, 7122 (2014).
15. Bhadra, C. M. *et al.* Antibacterial titanium nano-patterned arrays inspired by dragonfly wings. *Sci. Rep.* **5**, 16817 (2015).
16. Hasan, J., Jain, S. & Chatterjee, K. Nanoscale Topography on Black Titanium Imparts Multi-biofunctional Properties for Orthopedic Applications. *Sci. Rep.* **7**, 41118 (2017).
17. Cao, Y. *et al.* Nanostructured titanium surfaces exhibit recalcitrance towards Staphylococcus epidermidis biofilm formation. *Sci. Rep.* **8**, 1071 (2018).
18. Hizal, F. *et al.* Impact of 3D Hierarchical Nanostructures on the Antibacterial Efficacy of a Bacteria-Triggered Self-Defensive Antibiotic Coating. *ACS Appl. Mater. Interfaces* **7**, 20304–20313 (2015).
19. Wu, S., Zuber, F., Brugger, J., Maniura-Weber, K. & Ren, Q. Antibacterial Au nanostructured surfaces. *Nanoscale* **8**, 2620–2625 (2016).
20. Wu, S., Zuber, F., Maniura-Weber, K., Brugger, J. & Ren, Q. Nanostructured surface topographies have an effect on bactericidal activity. *J. Nanobiotechnology* **16**, 20 (2018).
21. Dickson, M. N., Liang, E. I., Rodriguez, L. A., Vollereaux, N. & Yee, A. F. Nanopatterned polymer surfaces with bactericidal properties. *Biointerphases* **10**, 021010 (2015).
22. Lu, X. *et al.* Silver carboxylate metal–organic frameworks with highly antibacterial activity and biocompatibility. *J. Inorg. Biochem.* **138**, 114–121 (2014).
23. Berchel, M. *et al.* A silver-based metal–organic framework material as a 'reservoir' of bactericidal metal ions. *New J. Chem.* **35**, 1000–1003 (2011).
24. Liu, Y. *et al.* Multiple topological isomerism of three-connected networks in silver-based metal–organoboron frameworks. *Chem. Commun.* **46**, 2608–2610 (2010).
25. Jaros, S. W. *et al.* Aliphatic Dicarboxylate Directed Assembly of Silver(I) 1,3,5-Triaza-7-phosphaadamantane Coordination Networks: Topological Versatility and Antimicrobial Activity. *Cryst. Growth Des.* **14**, 5408–5417 (2014).

26. Wang, X., Zhao, D., Tian, A. & Ying, J. Three 3D silver-bis(triazole) metal–organic frameworks stabilized by high-connected Wells–Dawson polyoxometallates. *Dalton Trans.* **43**, 5211–5220 (2014).
27. Abbasi, A. R., Akhbari, K. & Morsali, A. Dense coating of surface mounted CuBTC Metal–Organic Framework nanostructures on silk fibers, prepared by layer-by-layer method under ultrasound irradiation with antibacterial activity. *Ultrason. Sonochem.* **19**, 846–852 (2012).
28. Rodríguez, H. S., Hinestroza, J. P., Ochoa-Puentes, C., Sierra, C. A. & Soto, C. Y. Antibacterial activity against Escherichia coli of Cu-BTC (MOF-199) metal-organic framework immobilized onto cellulosic fibers. *J. Appl. Polym. Sci.* **131**, (2014).
29. Chiericatti, C., Basilico, J. C., Zapata Basilico, M. L. & Zamaro, J. M. Novel application of HKUST-1 metal–organic framework as antifungal: Biological tests and physicochemical characterizations. *Microporous Mesoporous Mater.* **162**, 60–63 (2012).
30. Arpa Sancet, M. P. *et al.* Surface anchored metal-organic frameworks as stimulus responsive antifouling coatings. *Biointerphases* **8**, 29 (2013).
31. Sheta, S. M., El-Sheikh, S. M. & Abd-Elzaher, M. M. Simple synthesis of novel copper metal–organic framework nanoparticles: biosensing and biological applications. *Dalton Trans.* **47**, 4847–4855 (2018).
32. Yu, Z. *et al.* Robust Chiral Metal–Organic Framework Coatings for Self-Activating and Sustainable Biofouling Mitigation. *Adv. Mater.* 2407409 (2024) doi:10.1002/adma.202407409.
33. Pezeshkpour, V. *et al.* Ultrasound assisted extraction of phenolic acids from *broccoli* vegetable and using sonochemistry for preparation of MOF-5 nanocubes: Comparative study based on micro-dilution broth and plate count method for synergism antibacterial effect. *Ultrason. Sonochem.* **40**, 1031–1038 (2018).
34. Colinas, I. R., Rojas-Andrade, M. D., Chakraborty, I. & Oliver, S. R. J. Two structurally diverse Zn-based coordination polymers with excellent antibacterial activity. *CrystEngComm* **20**, 3353–3362 (2018).
35. Kermanshahi, P. K. & Akhbari, K. The antibacterial activity of three zeolitic-imidazolate frameworks and zinc oxide nanoparticles derived from them. *RSC Adv.* **14**, 5601–5608 (2024).
36. Tamames-Tabar, C. *et al.* A Zn azelate MOF: combining antibacterial effect. *CrystEngComm* **17**, 456–462 (2014).
37. Zhuang, W. *et al.* Highly Potent Bactericidal Activity of Porous Metal-Organic Frameworks. *Adv. Healthc. Mater.* **1**, 225–238 (2012).
38. Aguado, S. *et al.* Antimicrobial activity of cobalt imidazolate metal–organic frameworks. *Chemosphere* **113**, 188–192 (2014).
39. Yuan, Y. & Zhang, Y. Enhanced biomimic bactericidal surfaces by coating with positively-charged ZIF nano-dagger arrays. *Nanomedicine Nanotechnol. Biol. Med.* **13**, 2199–2207 (2017).
40. Cheng, Y., Ma, X., Franklin, T., Yang, R. & Moraru, C. I. Mechano-Bactericidal Surfaces: Mechanisms, Nanofabrication, and Prospects for Food Applications. *Annu. Rev. Food Sci. Technol.* **14**, 449–472 (2023).
41. Linklater, D. P. *et al.* Mechano-bactericidal actions of nanostructured surfaces. *Nat. Rev. Microbiol.* **19**, 8–22 (2021).
42. Forgan, R. S. Modulated self-assembly of metal–organic frameworks. *Chem. Sci.* **11**, 4546–4562 (2020).
43. Shan, B. *et al.* Influences of Deprotonation and Modulation on Nucleation and Growth of UiO-66: Intergrowth and Orientation. *J. Phys. Chem. C* **122**, 2200–2206 (2018).
44. Sisican, K. M. D., Usman, K. A. S., Bacal, C. J. O., Edañol, Y. D. G. & Conato, M. T. Benzoic Acid Modulation of MIL-88B(Fe) Nanocrystals toward Tunable Synthesis of MOF-Based Fenton-like Degradation Catalysts. *Cryst. Growth Des.* **23**, 8509–8517 (2023).
45. Miyamoto, M., Kohmura, S., Iwatsuka, H., Oumi, Y. & Uemiya, S. In situ solvothermal growth of highly oriented Zr-based metal organic framework UiO-66 film with monocrystalline layer. *CrystEngComm* **17**, 3422–3425 (2015).
46. Sobhi, H., Matthews, M. E., Grandy, B., Masnovi, J. & Riga, A. T. Selecting polymers for medical devices based on thermal analytical methods. *J. Therm. Anal. Calorim.* **93**, 535–539 (2008).

47. Saini, M., Singh, Y., Arora, P., Arora, V. & Jain, K. Implant biomaterials: A comprehensive review. *World J. Clin. Cases* **3**, 52–57 (2015).
48. Kim, S.-N., Lee, Y.-R., Hong, S.-H., Jang, M.-S. & Ahn, W.-S. Pilot-scale synthesis of a zirconium-benzenedicarboxylate UiO-66 for CO2 adsorption and catalysis. *Catal. Today* **245**, 54–60 (2015).
49. Wu, Y. *et al.* Optimized scalable synthesis and granulation of MIL-88B(Fe) for efficient arsenate removal. *J. Environ. Chem. Eng.* **10**, 108556 (2022).
50. Fu, M. *et al.* Facile preparation of MIL-88B-Fe metal–organic framework with high peroxidase-like activity for colorimetric detection of hydrogen peroxide in milk and beer. *Appl. Phys. A* **127**, 928 (2021).
51. Yamashita, T. & Hayes, P. Analysis of XPS spectra of Fe2+ and Fe3+ ions in oxide materials. *Appl. Surf. Sci.* **254**, 2441–2449 (2008).
52. Li, Z. *et al.* 2-Methylimidazole-modulated UiO-66 as an effective photocatalyst to degrade Rhodamine B under visible light. *J. Mater. Sci.* **56**, 1577–1589 (2021).
53. Bondarenko, L. *et al.* Fenton reaction-driven pro-oxidant synergy of ascorbic acid and iron oxide nanoparticles in MIL-88B(Fe). *New J. Chem.* **48**, 10142–10160 (2024).
54. Zhu, R. *et al.* Fe-Based Metal Organic Frameworks (Fe-MOFs) for Bio-Related Applications. *Pharmaceutics* **15**, 1599 (2023).
55. Rahimi, S. *et al.* Automated Prediction of Bacterial Exclusion Areas on SEM Images of Graphene–Polymer Composites. *Nanomaterials* **13**, 1605 (2023).
56. Pandit, S. *et al.* Precontrolled Alignment of Graphite Nanoplatelets in Polymeric Composites Prevents Bacterial Attachment. *Small* **16**, 1904756 (2020).
57. Liu, Z. *et al.* Biocompatible mechano-bactericidal nanopatterned surfaces with salt-responsive bacterial release. *Acta Biomater.* **141**, 198–208 (2022).
58. Jiang, R. *et al.* Thermoresponsive Nanostructures: From Mechano-Bactericidal Action to Bacteria Release. *ACS Appl. Mater. Interfaces* **13**, 60865–60877 (2021).
59. Kim, H.-K., Baek, H. W., Park, H.-H. & Cho, Y.-S. Reusable mechano-bactericidal surface with echinoid-shaped hierarchical micro/nano-structure. *Colloids Surf. B Biointerfaces* **234**, 113729 (2024).
60. Kwon, O. *et al.* Computer-aided discovery of connected metal-organic frameworks. *Nat. Commun.* **10**, 3620 (2019).
61. Ghai, V. *et al.* Achieving Long-Range Arbitrary Uniform Alignment of Nanostructures in Magnetic Fields. *Adv. Funct. Mater.* 2406875 (2024) doi:10.1002/adfm.202406875.
62. Wang, X., Hou, J., Chen, F. & Meng, X. *In-situ* growth of metal-organic framework film on a polydopamine-modified flexible substrate for antibacterial and forward osmosis membranes. *Sep. Purif. Technol.* **236**, 116239 (2020).
63. Yu, Z. *et al.* Nacre-Inspired Metal-Organic Framework Coatings Reinforced by Multiscale Hierarchical Cross-linking for Integrated Antifouling and Anti-Microbial Corrosion. *Adv. Funct. Mater.* **33**, 2305995 (2023).
64. Harimawan, A., Rajasekar, A. & Ting, Y.-P. Bacteria attachment to surfaces – AFM force spectroscopy and physicochemical analyses. *J. Colloid Interface Sci.* **364**, 213–218 (2011).
65. Elbourne, A. *et al.* Bacterial-nanostructure interactions: The role of cell elasticity and adhesion forces. *J. Colloid Interface Sci.* **546**, 192–210 (2019).
66. Alam, F. & Balani, K. Adhesion force of *staphylococcus aureus* on various biomaterial surfaces. *J. Mech. Behav. Biomed. Mater.* **65**, 872–880 (2017).
67. Lee, I., Kim, J., Kwak, R. & Lee, J. Catalyst Droplet-Based Puncturable Nanostructures with Mechano-Bactericidal Properties Against Bioaerosols. *Adv. Funct. Mater.* **33**, 2213650 (2023).
68. Liu, T. *et al.* Mechanism Study of Bacteria Killed on Nanostructures. *J. Phys. Chem. B* **123**, 8686–8696 (2019).
69. Ivanova, E. P. *et al.* The multi-faceted mechano-bactericidal mechanism of nanostructured surfaces. *Proc. Natl. Acad. Sci.* **117**, 12598–12605 (2020).
70. Son, F. A. *et al.* Investigating the mechanical stability of flexible metal–organic frameworks. *Commun. Chem.* **6**, 1–8 (2023).
71. Cavka, J. H. *et al.* A New Zirconium Inorganic Building Brick Forming Metal Organic Frameworks with Exceptional Stability. *J. Am. Chem. Soc.* **130**, 13850–13851 (2008).

72. Wang, X.-G., Xu, L., Li, M.-J. & Zhang, X.-Z. Construction of Flexible-on-Rigid Hybrid-Phase Metal–Organic Frameworks for Controllable Multi-Drug Delivery. *Angew. Chem. Int. Ed.* **59**, 18078–18086 (2020).